\documentclass[twocolumn]{aastex61}

\usepackage{longtable}
\usepackage{graphicx}
\usepackage{amsmath,amssymb}
\usepackage{color}
\usepackage{units}
\usepackage{epstopdf}
\usepackage{hyperref}
\usepackage{multirow}
\usepackage{ifpdf}

\newcommand{\beq}{\begin{equation}}
\newcommand{\eeq}{\end{equation}}
\newcommand{\bdm}{\begin{displaymath}}
\newcommand{\edm}{\end{displaymath}}

\definecolor{Gray}{gray}{0.9}
\definecolor{orange}{rgb}{0.9,0.5,0}

\graphicspath{{./plots/}}

\begin{document}

\title{Towards rapid transient identification and characterization of kilonovae}
\author{Michael Coughlin}

\affil{Department of Physics, Harvard University, Cambridge, MA 02138, USA}

\author{Tim Dietrich}
\affil{Max Planck Institute for Gravitational Physics, Albert Einstein Institute, D-14476
Golm, Germany}

\author{Kyohei Kawaguchi}
\affil{Max Planck Institute for Gravitational Physics, Albert Einstein Institute, D-14476
Golm, Germany}

\author{Stephen Smartt}
\affil{Astrophysics Research Centre, School of Mathematics and Physics, 
       Queen's University Belfast, Belfast BT7 1NN, Northern Ireland UK}

\author{Christopher Stubbs}
\affil{Department of Physics, Harvard University, Cambridge, MA 02138, USA\\
Department of Astronomy, Harvard University, Cambridge MA 02138, USA}

\author{Maximiliano Ujevic}
\affil{Centro de Ciencias Naturais e Humanas, Universidade Federal do ABC, 
       09210-580, Santo Andŕe, Sao Paulo, Brazil}

\begin{abstract}
With the increasing sensitivity of advanced gravitational wave detectors, 
the first joint detection of an electromagnetic and 
gravitational wave signal from a compact binary merger will 
hopefully happen within this decade. 
However, current gravitational-wave likelihood sky areas span
$\sim 100-1000\,\textrm{deg}^2$, and thus it is a challenging task to identify 
which, if any, transient corresponds to the gravitational-wave event.
In this study, we make a comparison between recent kilonovae/macronovae 
lightcurve models for the purpose of assessing potential lightcurve templates 
for counterpart identification.
We show that recent analytical and parametrized models for these counterparts 
result in qualitative agreement with more complicated radiative transfer simulations.  
Our analysis suggests that with improved lightcurve models with smaller uncertainties,
it will become possible to extract information about ejecta properties and
binary parameters directly from the lightcurve measurement.
Even tighter constraints are obtained in cases for which gravitational-wave and kilonovae 
parameter estimation results are combined.
However, to be prepared for upcoming detections, more realistic kilonovae models are needed. 
These will require numerical relativity with more detailed microphysics, 
better radiative transfer simulations, and a better understanding of the underlying 
nuclear physics. 
\end{abstract}


\section{Introduction}
\label{sec:Intro}

The recent discovery of compact binary black hole systems 
\citep{AbEA2016a,AbEA2016g,AbEA2017} has initiated the era of 
gravitational-wave (GW) astronomy and even enhanced 
the interest in 
the combined observation of an electromagnetic (EM) and a GW signal 
\citep{AbEA2016b}. Currently, GW skymaps contain likelihood sky areas 
spanning $\approx 100-1000\,\textrm{deg}^2$ 
\citep{Fair2009,Fair2011,Grover:2013,WeCh2010,SiAy2014,SiPr2014,BeMa2015}; 
thus, it is essential to be able to differentiate transients associated 
with GW events from other transients. Models for potential EM 
emission from compact binary mergers remain highly uncertain, but 
emission timescales ranging from seconds to months and wavelengths from 
X-ray to radio can be expected \citep{Nakar2007,MeBe2012}.

Due to the large uncertainties in the sky localizations from the GW detectors, 
wide-field survey telescopes are needed to enable an EM follow-up study. 
Examples of current and future wide-field telescopes are the Panoramic 
Survey Telescope and Rapid Response System (Pan-STARRS) \citep{MoKa2012}, 
Asteroid Terrestrial-impact Last Alert System (ATLAS) \citep{Ton2011},
the intermediate Palomar Transient Factory (PTF) \citep{RaSh2009}, 
what will become the Zwicky Transient Facility (ZTF), and the Large 
Synoptic Survey Telescope (LSST) \citep{Ivezic2014}.

There are a variety of automatic schemes 
in surveys such as iPTF/ZTF \citep{MaCe2016} and Pan-STARRS \citep{SmCh2016}
trying to determine which transients are unassociated with the GW trigger.
For example, asteroids, variable stars, and active galactic nuclei are 
all objects that form the background for these searches, 
and, therefore, have to be removed \cite{CoBe2015}.
In general, background supernovae are the transients 
that remain after these cuts.
To further reduce the number of candidates, transients with host galaxies 
beyond the reach of the GW detectors are also removed. 
In addition, photometric evolution can be used to discriminate 
recent transients from old supernovae.
After spectra are taken, they are cross-matched against 
a library of supernovae, where they can be classified as 
Type Ia supernovae (SN Ia), two hydrogen-rich core-collapse 
supernovae  (SN  II), active galactic nuclei, etc.
The remaining transients which could not be identified 
might then be connected to the GW trigger. 

A variety of potential EM counterparts have been theorized to accompany the GW detection of a compact binary containing at least one neutron star, 
e.g.~short gamma ray bursts, kilonovae or radio burst. 
Among the most promising ``smoking guns'' of GW detections 
are kilonovae (also called macronovae) \cite{MeBe2012}. 
Kilonovae are produced during the merger of a binary neutron star (BNS) 
or a black hole-neutron star (BHNS) system. 
They last over a week, peak in the near-infrared with luminosities 
$\approx 10^{40}-10^{41}$\,ergs/s \citep{MeBa2015,BaKa2013} and 
are powered by the decay of radioactive r-process nuclei in 
the ejected material produced during the compact binary merger, 
see~\cite{Me2016,Ta2016} for recent reviews (see also~\cite{Ro2015}
for a review about multi-messenger astronomy). 
Some studies point out that the electromagnetic emissions 
similar to kilonovae can also be produced in the different mechanism~\cite{2014MNRAS.437L...6K,2015ApJ...802..119K}.
Material is ejected because of processes such as torque inside the tidal tails 
of the neutron stars, high thermal pressure produced by shocks created during 
the collision of two neutron stars, as well as neutrino or magnetic-field-driven winds. 
In reality, different ejecta mechanisms act simultaneously producing 
unbound material with complex morphology and composition. 

To model kilonovae properties as realistically as possible, 
full numerical relativity (NR) simulations and 
radiative transfer simulations have to be combined. 
NR simulations are needed to study the merger process and the 
different ejecta mechanisms. However, because those simulations 
only cover about a few hundred milliseconds around 
the compact binary merger, our knowledge about ejecta mechanisms 
acting on a longer timescale, due to magnetic field driven 
winds etc., is still limited, e.g.~\cite{SiMe2017}. 
Once the ejecta properties (ejecta mass, velocity, composition, morphology) 
are extracted from full NR simulations, this information can be used to set 
up radiative transfer simulations from which the 
lightcurve of the kilonova can be computed. 
However, because of the complexity of NR and radiative transfer simulations,
and due to our ignorance of astrophysical processes acting during 
the merger and postmerger of two compact objects, a variety of kilonovae 
approximants exist. 

In this paper, we shortly review some of the existing kilonovae models.
In particular, we will compare the parameterized models of~\cite{KaKy2016} and~\cite{DiUj2017}
against themselves and other kilonovae/macronovae models and radiative transfer 
simulations \citep{BaKa2016,RoFe2017,TaHo2014}. 
We ask the question of how much the models vary in their own parameters, 
using parameter estimation techniques to show plausible posteriors 
in case of a counterpart detection.
We will study how robust they are in terms of approximating other lightcurves 
and briefly compare the parameterized models to an example of a background 
contaminant, SN Ia using the SALT2 spectro-photometric empirical model \citep{GuAs2007}.
We explore the parameter degeneracies that arise from measurement 
of ejecta mass and velocity, $M_{\rm ej}$ and $v_{\rm ej}$, including the interplay 
between the measurement of masses and neutron star compactness.
We then consider the potential benefits of joint GW and EM parameter estimation.

\section{Motivation}
\label{sec:Motivation}

It is reasonable to question the purpose of parameter estimation of 
lightcurves with models which still might miss important astrophysical 
processes and which have systematic errors. Let us envision that we have 
a lightcurve from a transient consistent with both the time of the GW 
trigger and the skymap. There have been a number of cases where 
transients have been identified with these parameters, and it was 
necessary to determine their potential association with the GW event 
\citep{SmCh2016,SmCh2016b,SmCh2017}. 
In this way, there is a 
significant benefit to be able to show consistency between a measured 
lightcurve and an expected model to lend credibility to the association 
between the GW and EM trigger.\\
This is similar to the case of the 
first GW detection (which did not have an identified EM counterpart), where parameter estimation did not play a leading 
role in the assessment of the significance, but was important for 
verification that the detection was indeed real. 

Furthermore, for the ideal case in which a well-sampled lightcurve, 
mass posteriors from LIGO measurements, as well as a distance estimate 
from a host galaxy are available, we can use the distance from the host 
and convert apparent into absolute magnitudes. 
For such a case and with the availability of trustworthy models 
we do not need to allow for any zeropoint 
or time offset and would be able to place stringent constraints on the binary parameters
directly from the kilonova measurement. 

Finally, with significantly improved kilonovae models 
based on more accurate NR and radiative transfer 
simulations, including improved knowledge about nuclear physical properties, 
it might become possible to directly extract information of 
the compact binary from a well-sampled lightcurve from a kilonova counterpart measured in multiple bands, 
e.g.~by a telescope such as Pan-STARRS. 
This would allow for access to the properties of individual compact binary mergers even 
in the case where no GW signal or only a single detector trigger was present.

\section{Models}
\label{sec:Models}

\subsection{Kilonova Models}

As pointed out, to perform accurate NR and radiative transfer 
simulations remains a challenging task and further work including a 
better microphysical treatment is needed to allow a 
detailed understanding of ejecta, r-processes, and EM emission. 
However, in addition to the numerical work, a handful of analytical 
models have also been developed with the purpose of approximating kilonovae lightcurves. 
In the following, we give a brief overview about 
some approaches without guarantee of completeness. \\

One kilonovae model in which radioactively-powered transients are produced by 
accretion disk winds after the compact object merger was proposed 
by~\cite{KaFe2015}. 
In this model, the lightcurves contain two distinct components 
consisting of a $\approx$\,2 day blue optical transient and 
$\approx$\,10 day infrared transient. 
For this model, mergers resulting in a longer-lived neutron star or 
a more rapidly spinning black hole result in a brighter and bluer transient.

Another model driven by the merger of two neutron stars, 
where material ejected during or following the merger 
assembles into heavy elements by the r-process, is given in \cite{KaBa2013}. 
EM emission then occurs during the radioactive decay of the resulting nuclei.
\cite{BaKa2016} explore the emission profiles of the radioactive decay products, 
which include non-thermal $\beta $-particles, $\alpha $-particles, 
fission fragments, and $\gamma $-rays, 
and the efficiency with which their kinetic energy is absorbed by the ejecta. 
By determining the net thermalization efficiency for each particle type and 
implementing the results into detailed radiation transport simulations, 
they provide kilonova light curve predictions.
\cite{MeBa2015} also explore the $\beta$-decay of 
the ejecta mass powering a ``precursor'' to the main kilonova emission, 
which peaks on a timescale of a few hours in the blue.
\cite{RoFe2017} use semi-analytical models based on nuclear network simulations 
studying in detail the effect of the nuclear heating rate and 
ejecta electron fraction. The work of \cite{RoFe2017} shows in 
detail how lightcurve predictions change significantly for different 
nuclear physics parameters, e.g., the usage of different mass models. \\

Based on NR simulations, \cite{KaKy2016} derive fitting 
formulas for the mass and the velocity of ejecta from a generic BHNS merger 
and combine this with an analytic model of the kilonova 
lightcurve based on the radiative Monte-Carlo (MC) simulations of 
\cite{TaHo2014}.
\cite{DiUj2017} expand this work by using a large set of 
NR simulations to explore the EM signals from BNSs. 
The NR fit estimating the ejecta mass, velocity and morphology is extended 
by an analytical model also based on the radiative MC simulations of~\cite{TaHo2014}. 

Parametrized models as proposed in \cite{KaKy2016} and \cite{DiUj2017} 
directly tie GW parameters to expectations 
about the potential kilonova counterpart.
They do not require NR and radiative transfer simulations to be completed, 
which is an impossible task over the few days of observations.
Assumptions about the EOS of neutron stars, as well as measurement of 
the mass of the compact objects involved, allow 
the computation of the luminosity and lightcurves of kilonovae.

\subsection{Luminosity predictions}

\begin{figure*}[t]
 \includegraphics[width=3.5in]{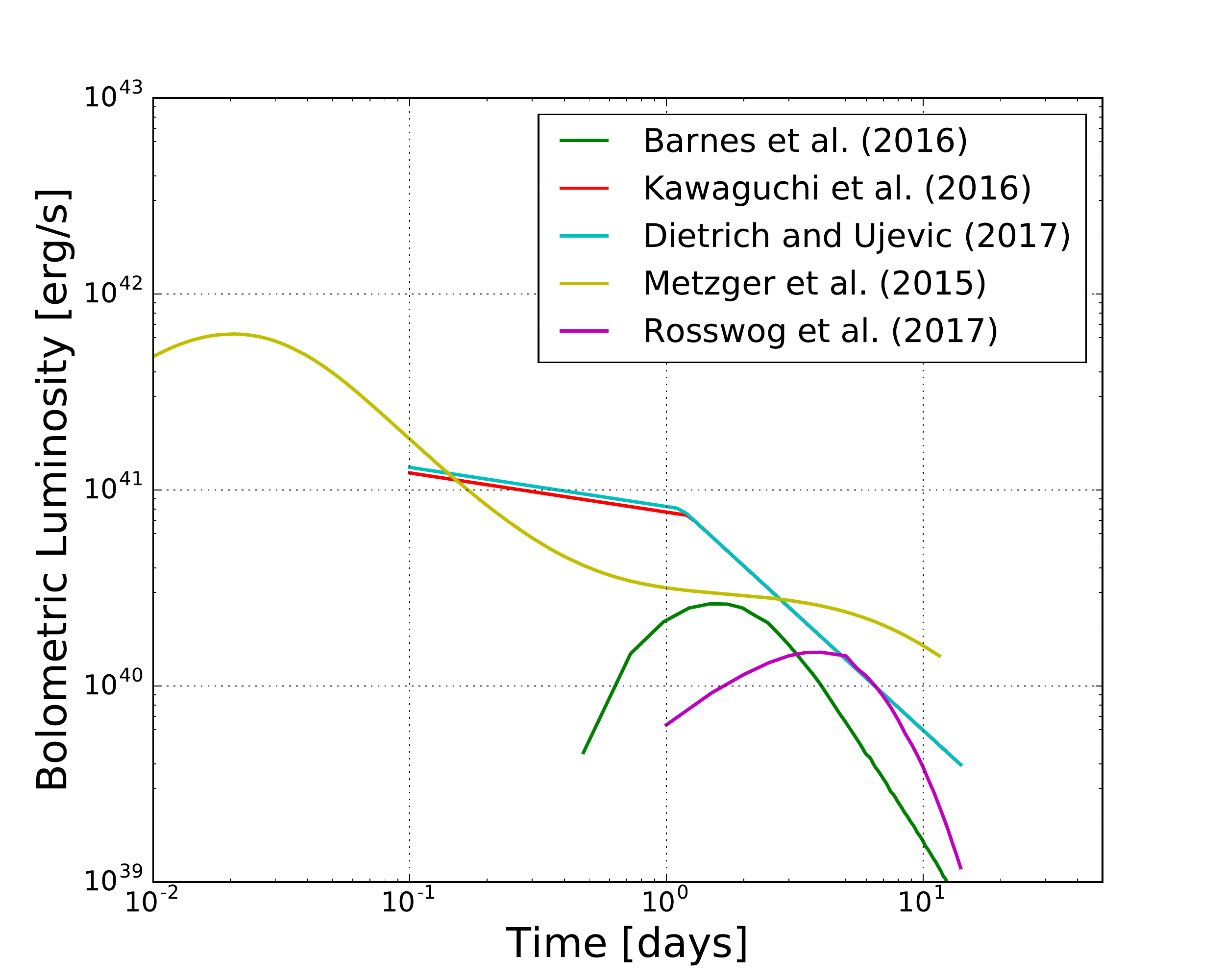}  
 \includegraphics[width=3.5in]{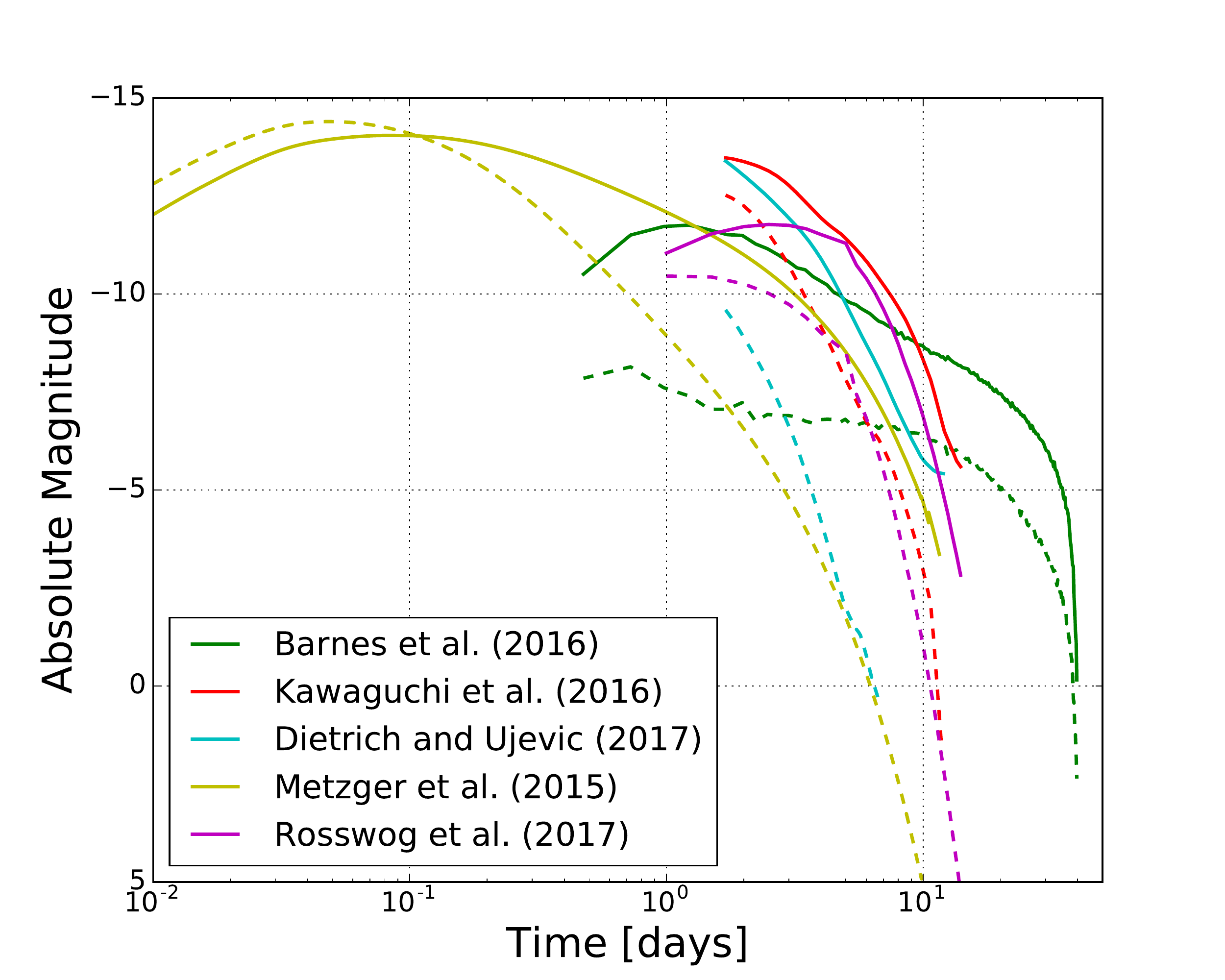}  
 \caption{
 We show the bolometric luminosity (left) and 
 the lightcurves in the $g$- (dashed) and $i$- (solid) bands (right).
 The parameterized models of \cite{KaKy2016} and  
 \cite{DiUj2017} use $M_{\rm ej} \approx 5 \times 10^{-3}$, 
 $v_{\rm ej} \approx 0.2$, and $\theta = 0.2$\,rad.
 \cite{BaKa2016} uses a model with $M_{\rm ej} \approx 5 \times 
10^{-3}$ and $v_{\rm ej} \approx 0.2$. 
We use the fiducial model of \cite{MeBa2015}, which uses a 
neutron mass cut  $m_{\rm n} = 10^{-4}$, opacity of 
$\kappa = 30$\,$\textrm{cm}^2 \textrm{g}^{-1}$, and electron fraction $Y_{\rm e} = 0.05$. 
From \cite{RoFe2017}, we include a model with $M_{\rm ej} = 0.0079$
and $v_{\rm ej} = 0.12$ which is the closest available 
to our fiducial model.}
\label{fig:models}
\end{figure*}

Because the ejecta morphology, the thermalization efficiency, and the 
opacity are not well constrained, it is advantageous to use a variety of 
models that estimate these quantities in different ways. In general, the 
luminosity will depend on the thickness of the ejecta, which is one of 
the main differences between BNS and BHNS systems. The thinner the 
ejecta becomes, the higher the density and temperature become. This 
affects the color temperature of the spectrum and consequently has a 
large impact on the detected lightcurve.

There are two limiting cases, (i) the ejecta are geometrically thick and 
approximately spherical and (ii) the ejecta are geometrically thin. In 
general, due to shock driven ejecta, BNS mergers correspond mostly to 
the former and BHNS systems to the latter case, however, a clear 
distinction is impossible. The morphology of ejecta affects the 
diffusion time scale and change the evolution of the lightcurve before 
the system becomes optically thin. When the system is optically thin, 
the difference in morphology may not be important for the lightcurve 
evolution anymore. Since information about ejecta velocity is primarily 
contained in the lightcurve during the optically thick phase, modeling 
of this phase is important to constraint ejecta velocity.\\

As a first comparison between different models, we consider a spherical 
ejecta with $M_{\rm ej} \approx 5 \times 10^{-3}$ and $v_{\rm ej} 
\approx 0.2$, see \cite{BaKa2016}. Here and in the following, we will 
give $v_{\rm ej}$ in fractions of the speed of light and masses in 
fractions of the mass of the sun $M_{\odot}$. For the non-spherical 
parametrized models of \cite{KaKy2016} and \cite{DiUj2017}, we further 
assume $\theta = 0.2$\,rad. From \cite{RoFe2017}, we include a model 
with $M_{\rm ej} = 0.0079$ and $v_{\rm ej} = 0.12$, which is closest to 
our fiducial model.

Additionally, we include the approximant of \cite{MeBa2015}, which 
focused on the blue transient produced at a time around merger, which 
uses a neutron mass cut $m_{\rm n} = 10^{-4}$, opacity of $\kappa = 
30$\,$\textrm{cm}^2 \textrm{g}^{-1}$, and electron fraction $Y_{\rm e} = 
0.05$.

Figure~\ref{fig:models} shows the bolometric luminosity and the 
lightcurves in the $g$- (dashed) and $i$- (solid) bands. The kilonovae 
models have significant short-term dynamics, with changes of more than a 
magnitude in less than a day. Both the \cite{KaKy2016} and 
\cite{DiUj2017} models are based on the MC simulations of 
\cite{TaHo2014} for which a constant thermal efficiency is assumed 
($\epsilon_{\rm th}=0.5$).

The model of \cite{BaKa2016} includes a time dependent efficiency, 
which leads to a faster decay of the bolometric luminosity and 
magnitude because after a few days after the merger the thermalization efficiency drops below the constant 
thermalization efficiency employed in the \cite{TaHo2014} simulations.
\cite{RoFe2017} employ both time dependent and constant efficiencies 
and use a more complex density profile. The model picked from \cite{RoFe2017} 
shows a smaller bolometric luminosity than other models, notice, however, that 
as shown in \cite{RoFe2017} the usage of different mass models effects the 
luminosity by about $\approx 600\%$, i.e., all presented models
come with large uncertainties and crucially depent on nuclear physics assumptions. 
The model of \cite{MeBa2015} describes 
the blue transient arising from a small fraction of the ejected mass 
which expands sufficiently rapidly such that the neutrons are not 
captured and instead $\beta$-decay, giving rise to a clear peak 
in the bolometric luminosity visible around the time of merger. 

Comparing \cite{DiUj2017} and \cite{KaKy2016} 
we see a clear difference in the $g$-band. This has already 
been pointed out in \cite{TaHo2014}.
The main difference seems to arise from the difference of 
employed bolometric corrections, 
which itself will depend on the ejecta morphology. 
Since BHNS ejecta are much more non-spherical 
and are concentrated in the equatorial plane, 
they have higher temperatures which make 
the spectrum bluer than BNS ejecta with the same mass. \\

\begin{figure}[t]
 \includegraphics[width=3.5in]{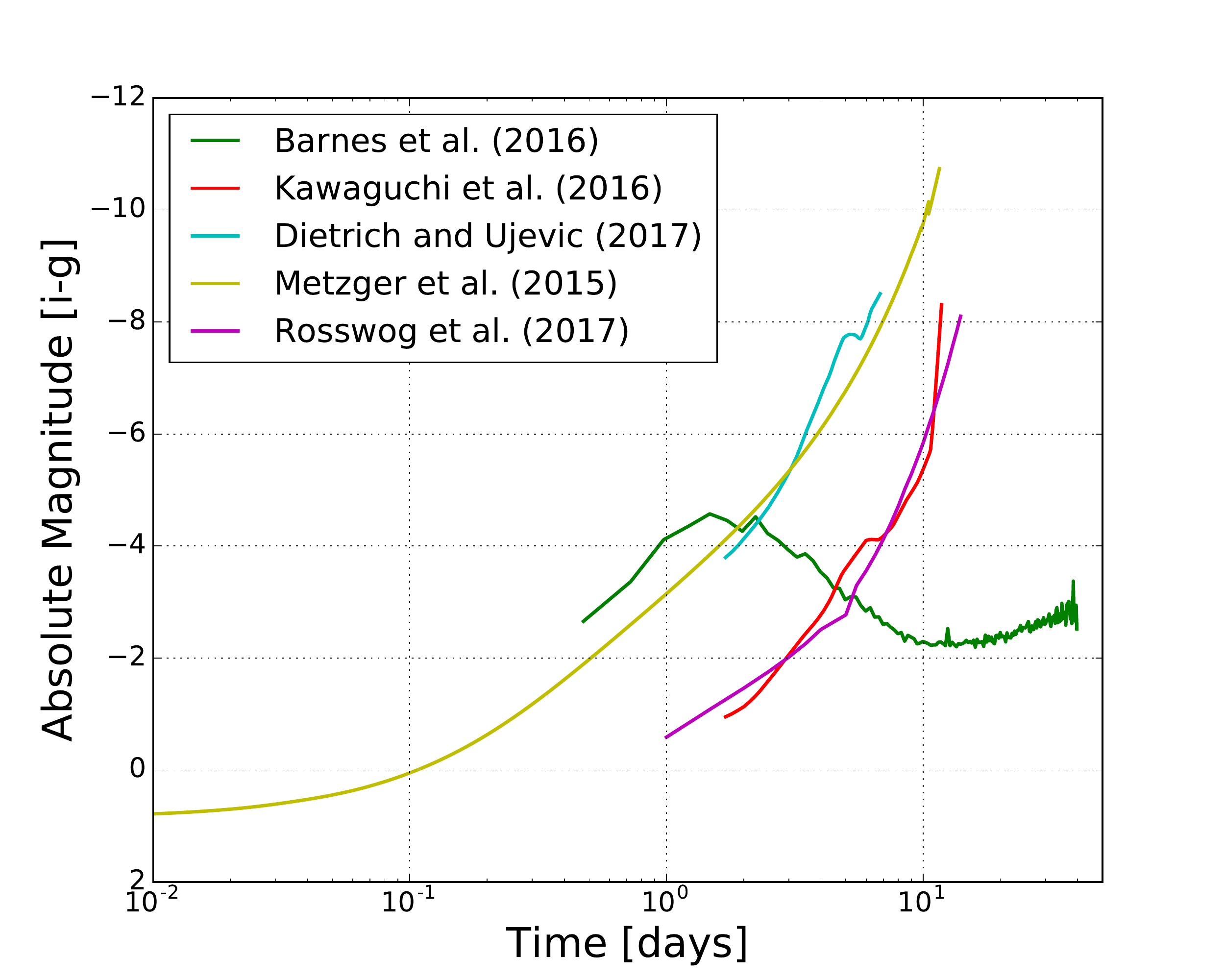}  
 \caption{
 Difference between the $g$- and $i$-bands for 
 the models presented in Figure~\ref{fig:models}.
 }
\label{fig:models_iminusg}
\end{figure}

We can take the opportunity of having a variety of kilonova models accessible to compare the lightcurve colors. It is common in dedicated 
searches for kilonovae to make color cuts \citep{DoKe2017}. Figure~\ref{fig:models_iminusg} shows the difference between the $g$- and 
$i$-bands for the models presented in Figure~\ref{fig:models}. As expected, all of the kilonova models show differences of at least 2\,mag, 
especially on later time scales. For this reason, independent of the employed kilonova model, the proposed analysis will optimize the strategy 
for the detection of GW optical counterparts. Given the relative consistency in color among the models, imaging the transients in both the 
blue/green and the near-infrared can help differentiate from other transients. Due to the high opacities of r-process nuclei, most models 
predict emission in the near infrared wavelengths. These observations are required within the first few days due to the faint magnitudes 
involved. As explained above, the significant changes in magnitude over day time-scales can also help differentiate them as compared to 
possible background transients such as SN Ia. 

\subsection{Dependence of the bolometric lightcurve on the density profile,
            morphology, and thermal efficiency}

\begin{figure}[t]
\includegraphics[width=3.5in]{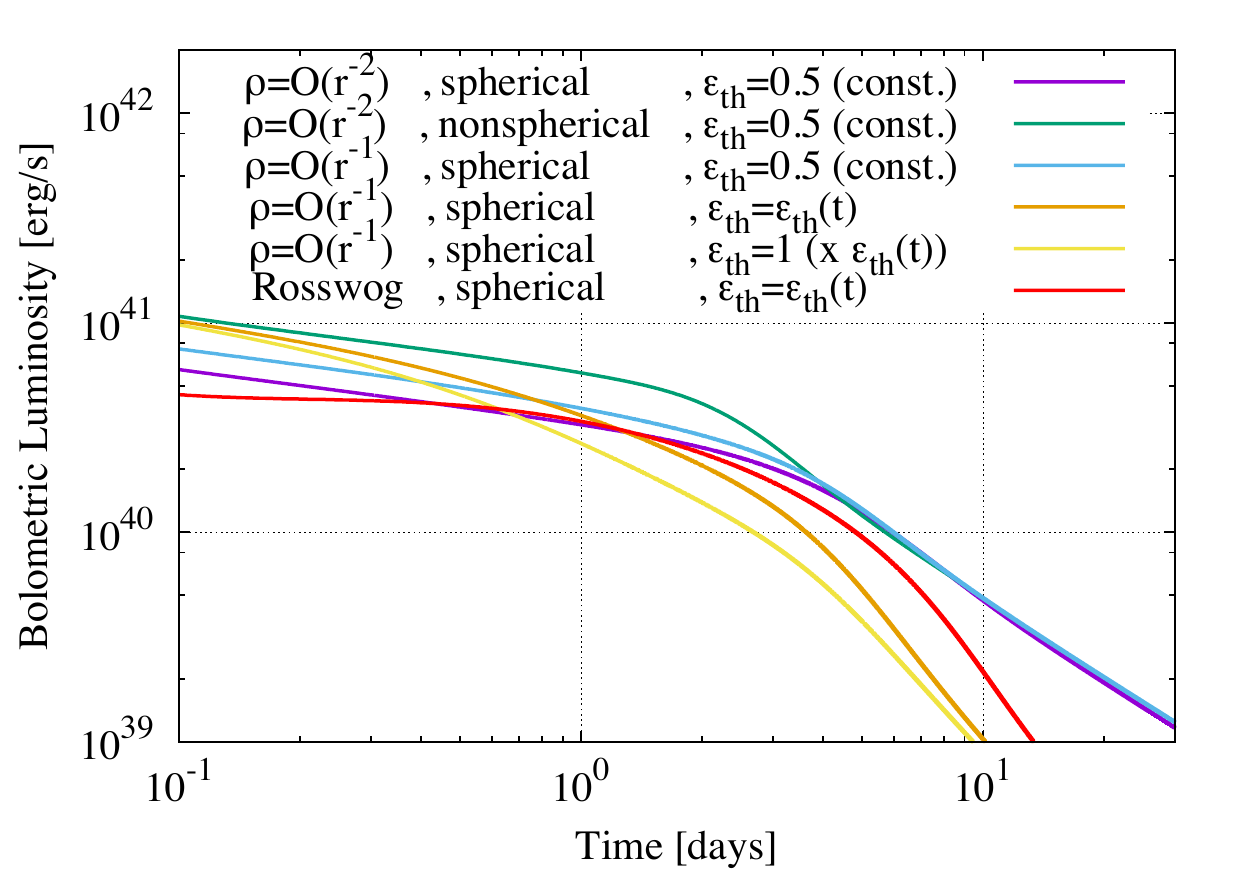}  
\caption{
The comparison of the bolometric luminosity for various setups: 
The purple line employs a $\rho\propto r^{-2}$ density profile, 
spherical geometry, and a constant thermalization efficiency 
($\epsilon_{\rm th} \approx 0.5$). 
The green line is similar to the purple curve but with non-spherical ejecta 
with $\theta_{\rm ej}=0.2$ and $\varphi=\pi$ 
[the same morphology employed in \cite{KaKy2016}, 
\cite{DiUj2017}, and 
\cite{BaKa2016}]. 
The blue curve is similar to the purple curve but with 
$\rho\propto r^{-1}$ density profile. The orange line is similar to 
the blue one but with time-(mass)-dependent 
thermalization efficiency of \cite{BaKa2016}. 
The yellow curve is similar to the blue curve but we employed a constant 
thermalization efficiency ($\epsilon_{\rm th}=1$) and multiplied afterwards by the 
time-(mass)-dependent thermalization efficiency given in \cite{BaKa2016}. 
The red curve denotes the 
bolometric lightcurve employing $M_{\rm ej}=0.0079$ and $v_{\rm ej}=0.12$, 
and the same density profile as in \cite{RoFe2017}.}
\label{fig:blc_comp}
\end{figure}

As shown in Figure~\ref{fig:models} (left panel), 
the bolometric luminosity of the models from 
\cite{KaKy2016}, \cite{DiUj2017}, 
\cite{BaKa2016}, and \cite{RoFe2017} 
can be significantly different (we do not include the blue transient 
proposed in \cite{MeBa2015} in the following analysis 
since it is powered by a different mechanism). 
While similar ejecta masses, velocities, and energy deposition rates are employed, 
the models use different density profiles, morphology, and thermalization efficiency. 
The models of \cite{KaKy2016} and \cite{DiUj2017} 
assume $\rho\propto r^{-2}$ for the density profile, 
non-spherical geometry, and a constant thermalization efficiency 
($\epsilon_{\rm th} \approx0.5$). 
The model of \cite{BaKa2016} assumes spherical ejecta 
with $\rho\propto r^{-1}$ for the density profile and the 
time-(mass)-dependent thermalization efficiency is taken into account.  
The model of \cite{RoFe2017} also assumes spherical ejecta with 
an homogeneously expanding density profile and
time-(mass)-dependent thermalization efficiency with the FRDM model.

To check how these differences affect the bolometric lightcurves, we 
perform a simple radiation transfer simulation varying the density 
profile, ejecta morphology, and thermalization efficiency. In this 
calculation, we assume the flux limited diffusion approximation of the 
radiative transfer \citep{LePo1981}, a constant gray opacity with 
$10\,{\rm cm^2 g^{-1}}$, and the heating rate that is employed in 
\cite{KaKy2016} and \cite{DiUj2017}.

Figure~\ref{fig:blc_comp} compares the bolometric luminosity 
for various setups. The figure clearly shows that different 
ejecta morphologies and thermalization efficiencies change 
the bolometric luminosity by about a factor of $\approx2$. 
This explains qualitatively the difference 
in the bolometric luminosity and lightcurves in 
Figure~\ref{fig:models}. The difference in the model of 
\cite{RoFe2017} is also explained by the difference in the 
ejecta mass, ejecta velocity, and thermalization efficiencies.
On the other hand, a different density profile 
has only a minor effect. 

These results indicate that for future development of analytical 
kilonovae approximants the focus should be put on modeling the ejecta 
morphology and the time-dependent thermalization efficiency. We also 
find that considering a constant thermalization efficiency of 
$\epsilon_{\rm th} = 1$ and then multiplying with $\epsilon_{\rm th}(t)$ 
(given in \cite{BaKa2016}) or directly employing a time dependent 
thermal efficiency leads only to differences of $\approx 40\%$. This 
suggests that, at least for the bolometric luminosity, the 
time-dependency of the thermalization efficiency can be approximately 
taken into account just by multiplying its function to the luminosity 
obtained by the constant efficiency. This is of particular importance 
for further improvement of the parametrized models which, at the current 
stage, are based on simulations employing a constant thermalization 
efficiency.\\

In addition to the discussed effects further uncertainties exist, which 
make a modeling and prediction of kilonovae luminosities difficult. 
\cite{RoFe2017} point out that the electron fraction and heating rate 
are main uncertainties in the 
current modeling of kilonovae lightcurves. They find that by using two different mass 
models (\cite{DuZu1995} (DZ31) and Finite Range Droplet Model~\citep{MoNi1993}) 
the bolometric luminosity can be different up to 
$\approx 600\%$. This is caused by the fact that the nuclear heating rate enters linearly 
into the bolometric luminosity. 

\section{Model Comparisons and Parameter Estimation}
\label{sec:PE}

In this section, we perform parameter estimation and model comparisons. 
We will use the \cite{KaKy2016} and \cite{DiUj2017} models to compare 
both to other models and against themselves. As described, there are two 
parts to each of these models: the ejecta fitting formulas and the 
kilonovae lightcurves. Avoiding the ejecta fitting formulas, we can 
improve efficiency and accuracy by directly sampling the ejecta mass 
$M_{\rm ej}$ and velocity $v_{\rm ej}$, and later employ the 
correlations between the ejecta mass properties, e.g.~ejecta mass 
$M_{\rm ej}$ and velocity $v_{\rm ej}$, and the binary parameters (see 
Section~\ref{extracting}). Furthermore, we sample over the latitudinal 
and longitudinal opening angles, denoted as $\theta_{\rm ej}$ and 
$\phi_{\rm ej}$, respectively. Opacity, $\kappa=10$\,$\textrm{cm}^2 
\textrm{g}^{-1}$, 
heating rate coefficient $\epsilon_0 = 1.58 \times 10^{10}$\,erg 
g$^{-1}$ s$^{-1}$, 
heating rate $\alpha = 1.2$, and thermalization efficiency 
$\epsilon_{\rm th} = 0.5$ are held fixed.

In this analysis, we will use a version of 
Multinest \citep{FeHo2009} commonly used in GW data 
analysis \citep{FeGa2009}, and wrapped in python \citep{BuGe2014}. 
This algorithm has the benefit of computing the Bayesian evidence 
for a given set of parameters, which can be used to assign relative probabilities 
to different models. The likelihood evaluation proceeds as follows. 
For each parameter set sampled, lightcurves in $griz$ bands are computed. 
We use linear extrapolation of the magnitudes to extend the lightcurves in cases where the 
model does not predict the full time covered by the target lightcurve.
In addition to the parameters above, we also allow the lightcurves 
to shift in time by an offset $T_0$, which allows for a measurement of the initial 
time of the kilonovae and therefore gives important evidence for a potential counterpart, 
and in magnitude by a color-independent zeropoint offset $\textrm{ZP}$, 
which compensates for our ignorance about the distance to the source. 
A $\chi^2$ distribution is then calculated between the lightcurve produced from 
the model and the target lightcurve. 
The likelihood is then simply that from a $\chi^2$ distribution. 
The priors used in the analyses are as follows:
$-5 \leq T_0 \leq 5$\,days, $-50.0 \leq \textrm{ZP} \leq 50.0$\,mag, $-5 
\leq \log_{10} (M_{\rm ej}) \leq 0$, $ 0 \leq v_{\rm ej} \leq 1$, $0 
\leq 
\theta \leq \pi/2$\,rad, and $0 \leq \phi \leq 2 \pi$\,rad.
The priors are flat over the stated ranges.

\subsection{Self-consistency check of parametrized models}

\begin{figure*}[h!]
 \includegraphics[width=3.5in]{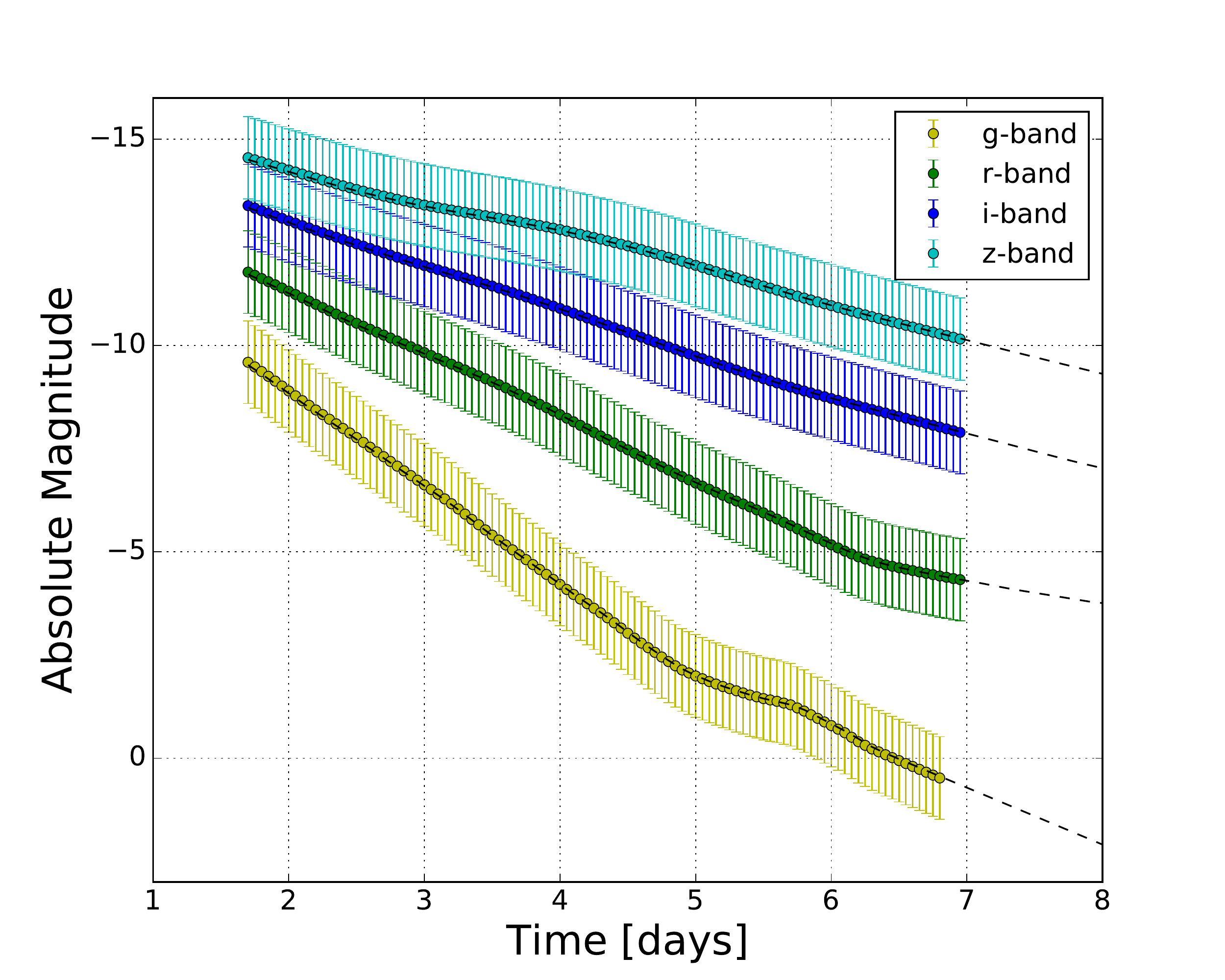}
 \includegraphics[width=3.5in]{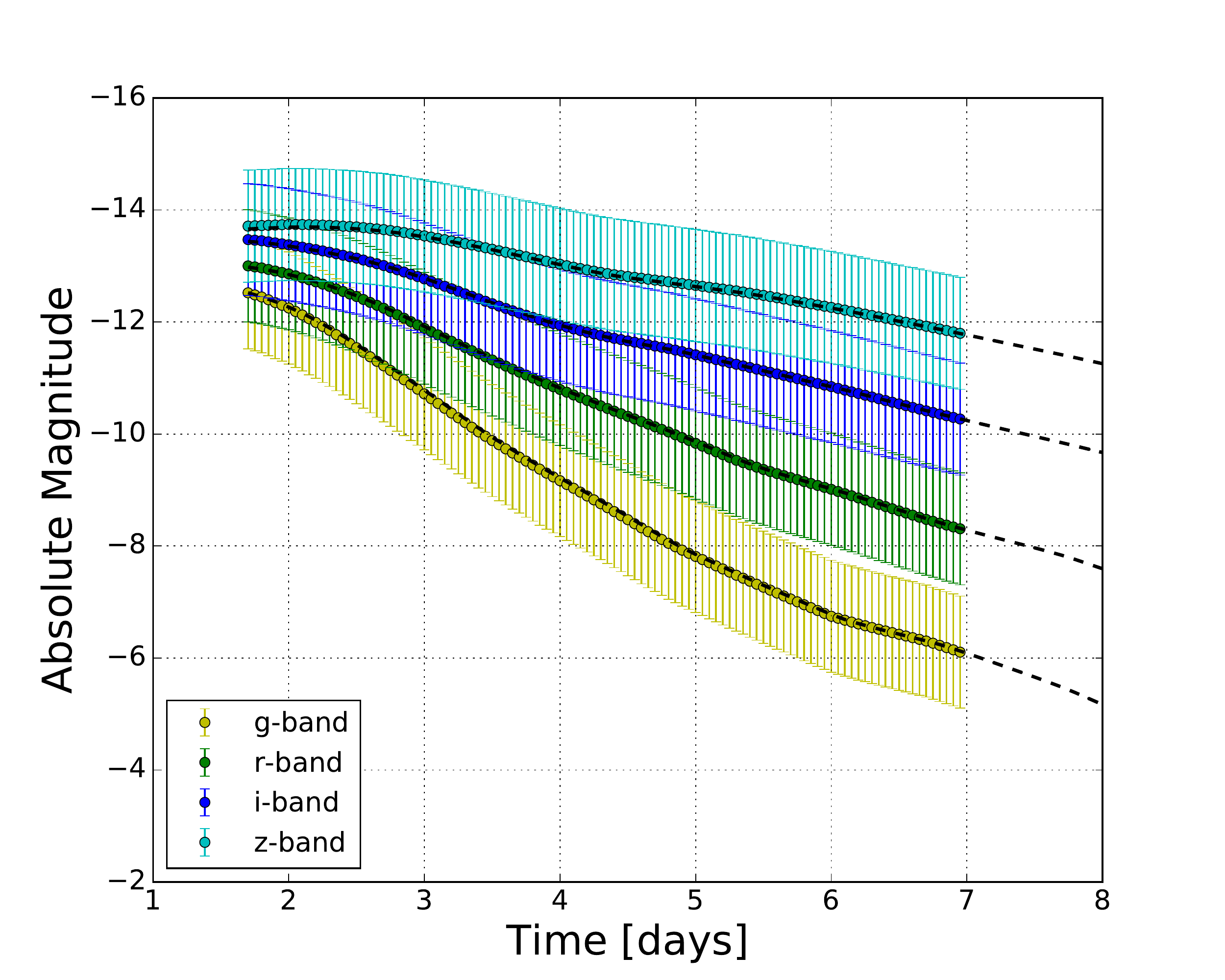} 
 \includegraphics[width=3.5in]{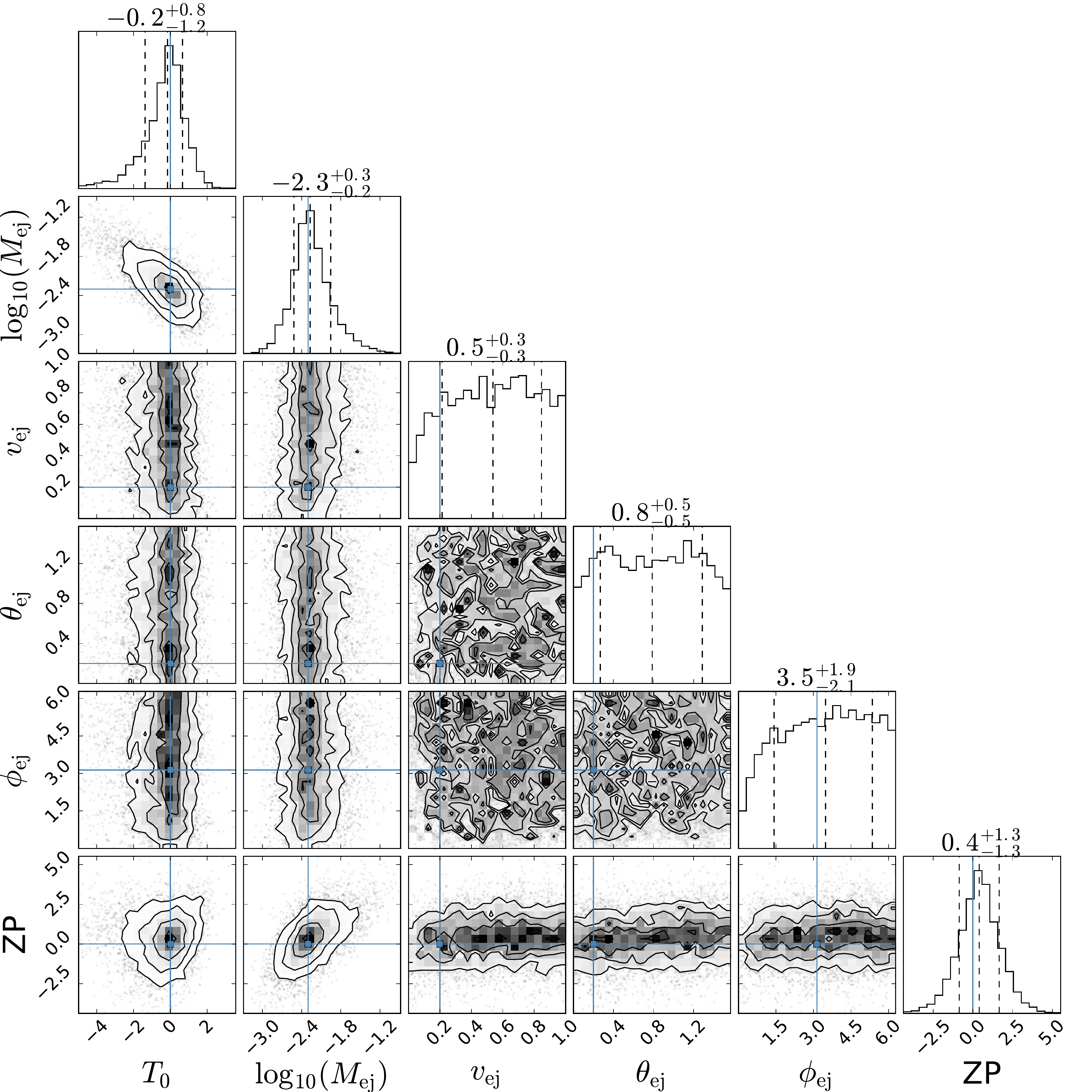}
 \includegraphics[width=3.5in]{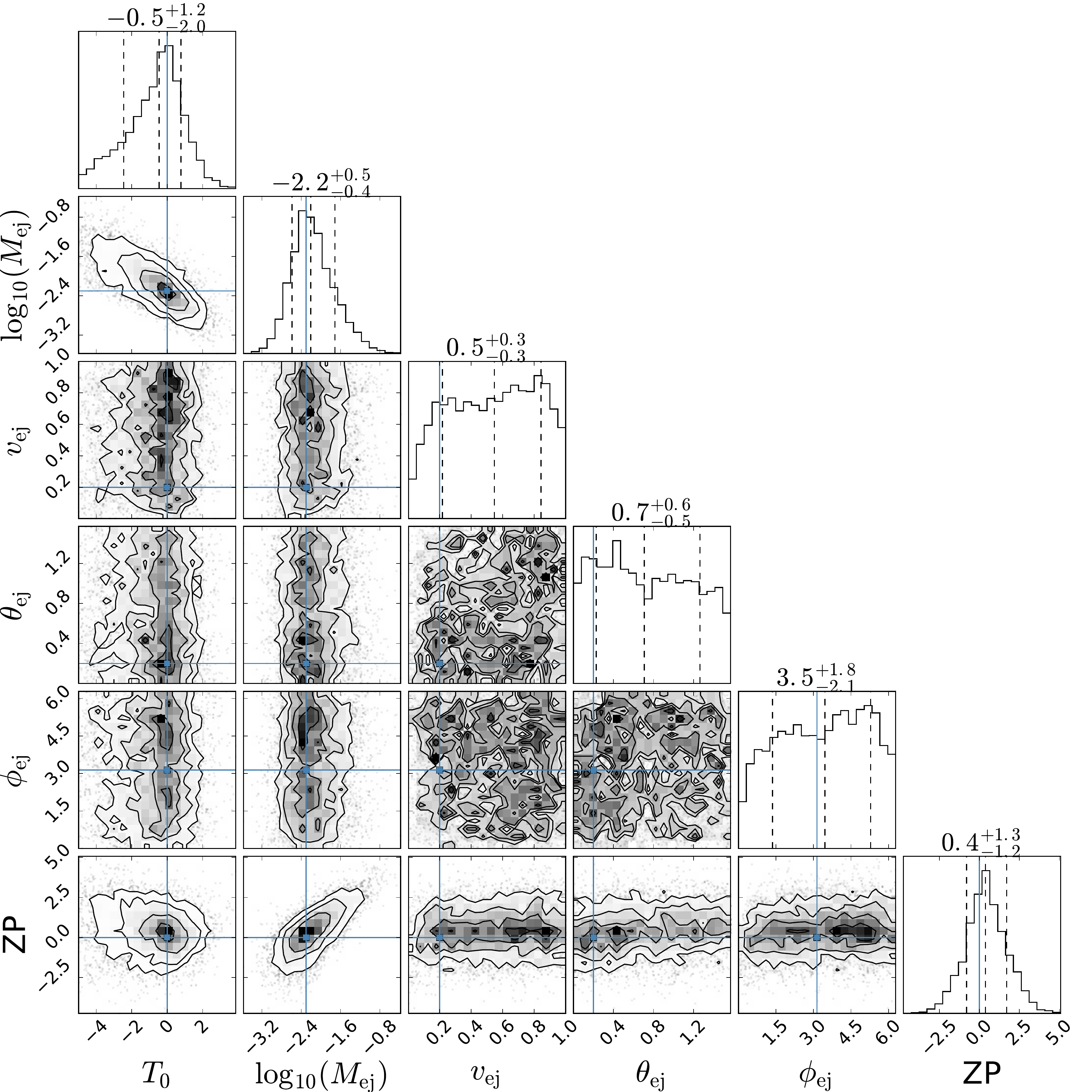} 
  \caption{
   The top row shows lightcurves for \cite{DiUj2017} (left) and 
   \cite{KaKy2016} (right). 
   We use lightcurves with $M_{\rm ej} = 5 \times 10^{-3}$, 
   $v_{\rm ej} = 0.2$, $\theta_{\rm ej} = 0.2$\,rad, and 
   $\phi_{\rm ej} = 3.14$\,rad for the lightcurve computation.   
   We also perform a maximum likelihood $\chi^2$ fit to 
   each lightcurve using the same models for comparison.
   The lines with error bars show the injected lightcurve with the assumed 1\,mag error budget.
   The dashed black lines show the best fit lightcurve to that model, including the linear extrapolation.      
   The bottom row shows the corresponding corner plots.
 }
 \label{fig:BNS_BHNS_fits}
\end{figure*}
 
\begin{figure*}[t]
 \includegraphics[width=3.5in]{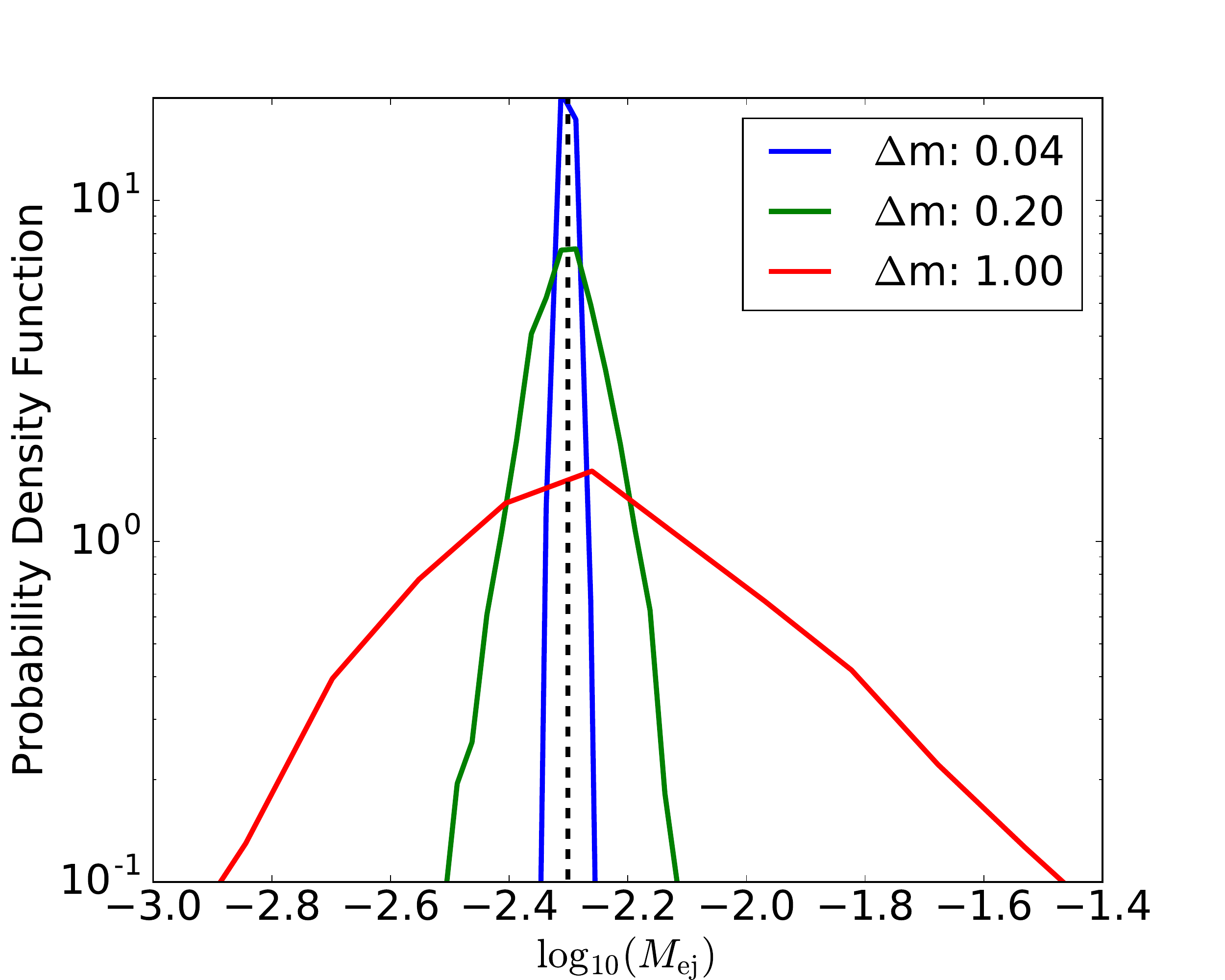}
 \includegraphics[width=3.5in]{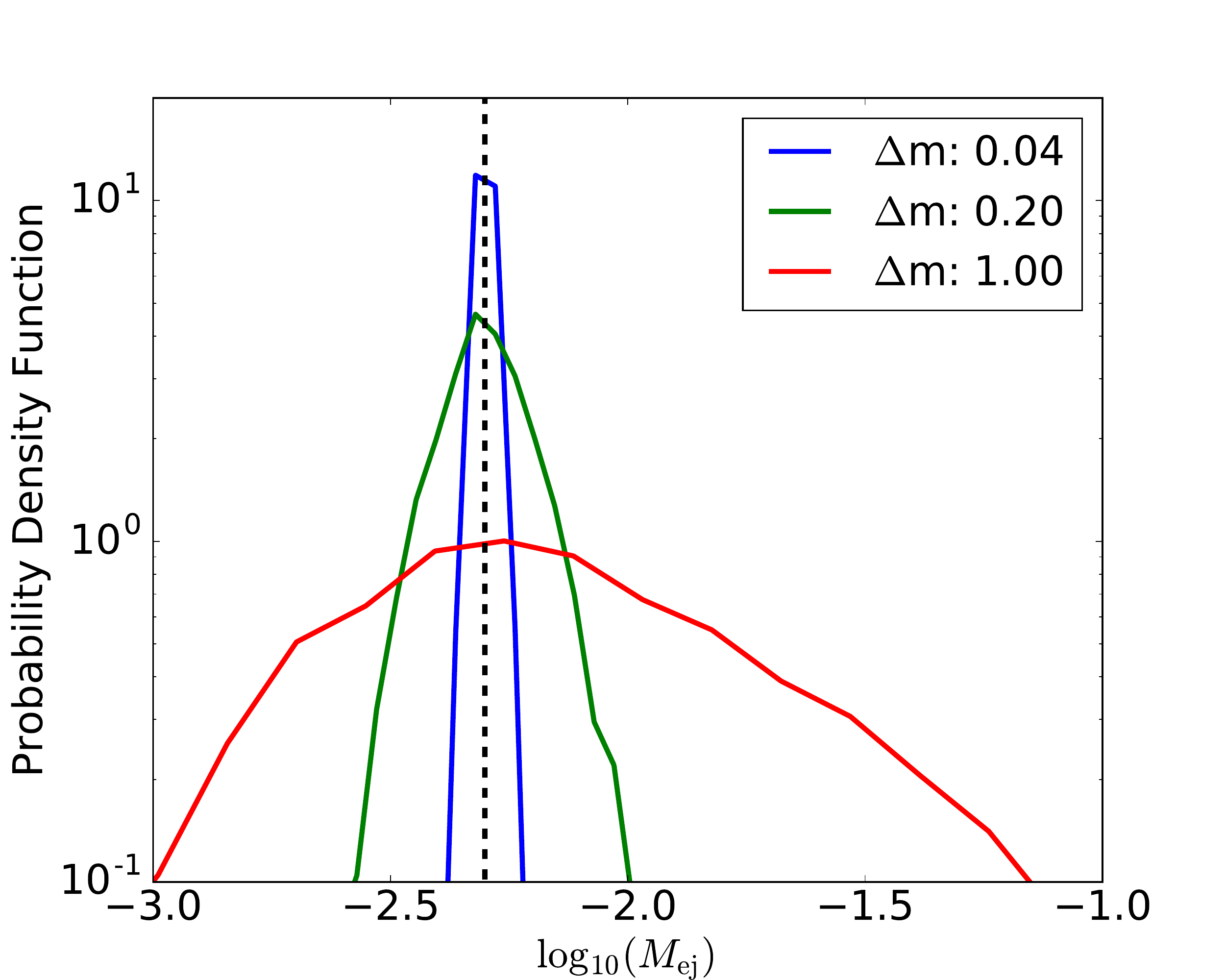} 
  \caption{
   Histograms of $M_{\rm ej}$ recovery. We inject lightcurves computed with the 
   parametrized model of \cite{DiUj2017} (left) and \cite{KaKy2016} 
   (right). 
   We recover the injected lightcurve with the same model. 
   For decreasing uncertainties assigned to the models, 
   the ejecta mass gets better constrained and approaches the true value (vertical dash-dotted).
 }
 \label{fig:BNS_BHNS_error_mej}
\end{figure*}

As a first test of the numerical method, 
we use lightcurves produced by the parametrized models 
for BNS and BHNS and also recover the ejecta properties with 
the same models. 
The top row of Figure~\ref{fig:BNS_BHNS_fits} 
shows lightcurves of such a comparison, where we  
assume uncertainties of the models of 1\,mag as stated 
in \cite{DiUj2017} and \cite{KaKy2016}. 
The top row shows that the injected lightcurves are recovered properly.
We quantify the level of overlap between parameters with 
``corner'' plots \citep{For2016}, shown in the bottom row of Figure~\ref{fig:BNS_BHNS_fits}.  
Shown are 1- and 2-D posteriors marginalized over the rest of the parameters.
In general, there are a few key features. 
First, with the $\approx 1$\,mag uncertainty associated to these models, 
a large number of lightcurves computed with the parametrized models 
are consistent with the injected/baseline lightcurve we took. 
This means that no strict parameter constraints can be obtained. 
But, although the models have stated $\approx 1$\,mag uncertainty, 
we can study a possible scenario with models having smaller uncertainties, 
e.g.~$\approx 0.2$\,mag or even $0.04$\,mag, which approximate the 
characteristic uncertainty for observations. 
In Figure~\ref{fig:BNS_BHNS_error_mej}, 
we show histograms for $M_{\rm ej}$ for the case where the uncertainties are varied. 
The figure demonstrates that $M_{\rm ej}$ constraints are significantly 
improved when the assigned error to the model is small. 
In particular for an uncertainty of $0.2$\,mag the ejecta mass can be determined 
up to $\log_{10} M_{\rm ej}\approx \pm 0.5$, and in cases where the 
uncertainty would be limited by the observation (uncertainty of $0.04$\,mag)
the ejecta mass could be determined to $\log_{10} M_{\rm ej}\approx \pm 0.1$.
This motivates the need for further improved parametrized models of 
kilonovae lightcurves. 

In contrast to the ejecta mass, the ejecta velocity is poorly constrained in our analysis.
This is because the analytic models do not include times $t\lesssim 1\,{\rm day}$, where the ejecta
are optically thick. However, 
the dependence on the ejecta velocity is only significant during this stage. Afterwards,
the lightcurves is primarily determined by the ejecta mass. 
Therefore, to improve the estimation of the ejecta velocity, 
extension of the lightcurve models to earlier times is required. 
 
\subsection{Comparison with Tanaka et al.}
 
\begin{figure*}[t]
 \includegraphics[width=3.5in]{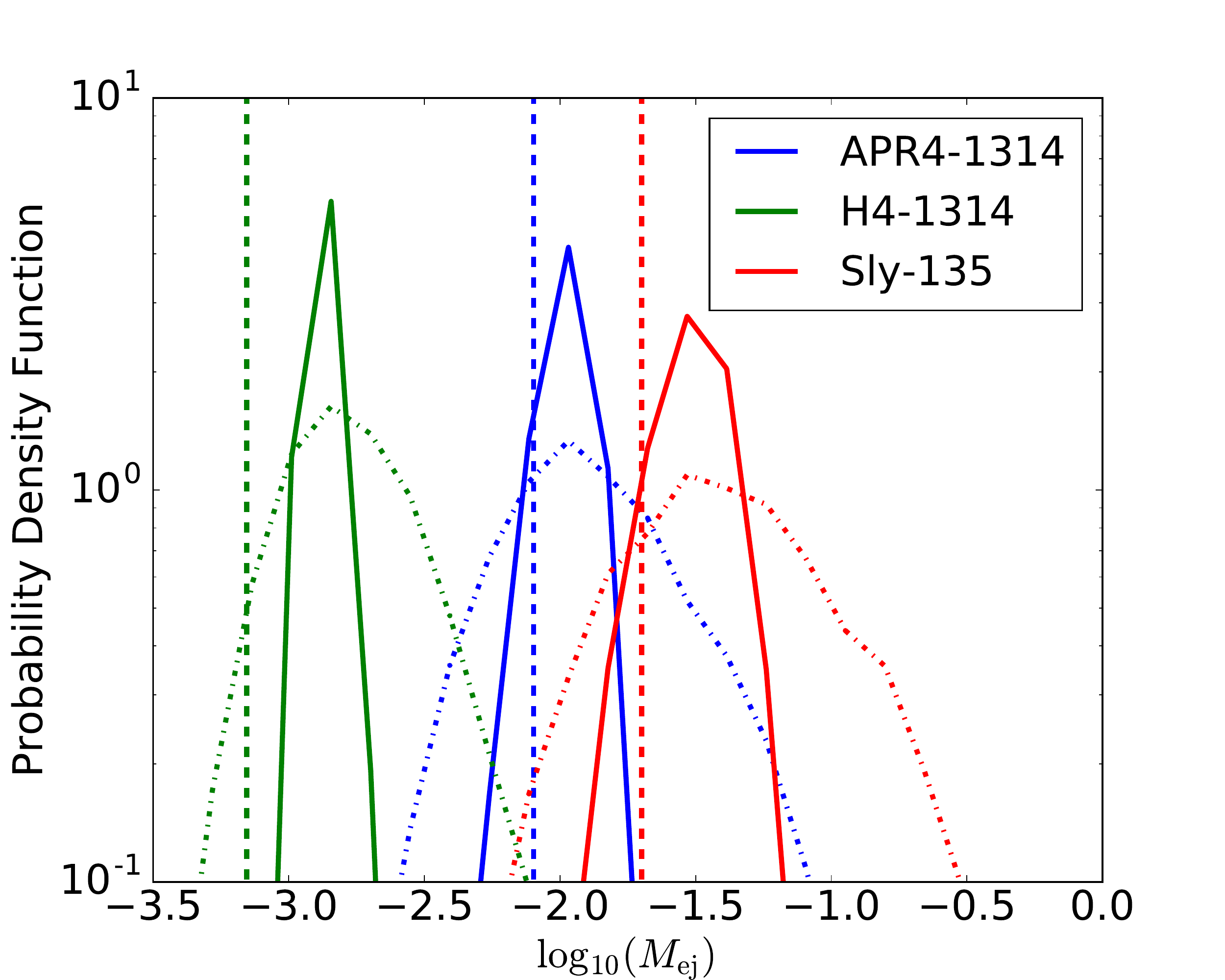}
 \includegraphics[width=3.5in]{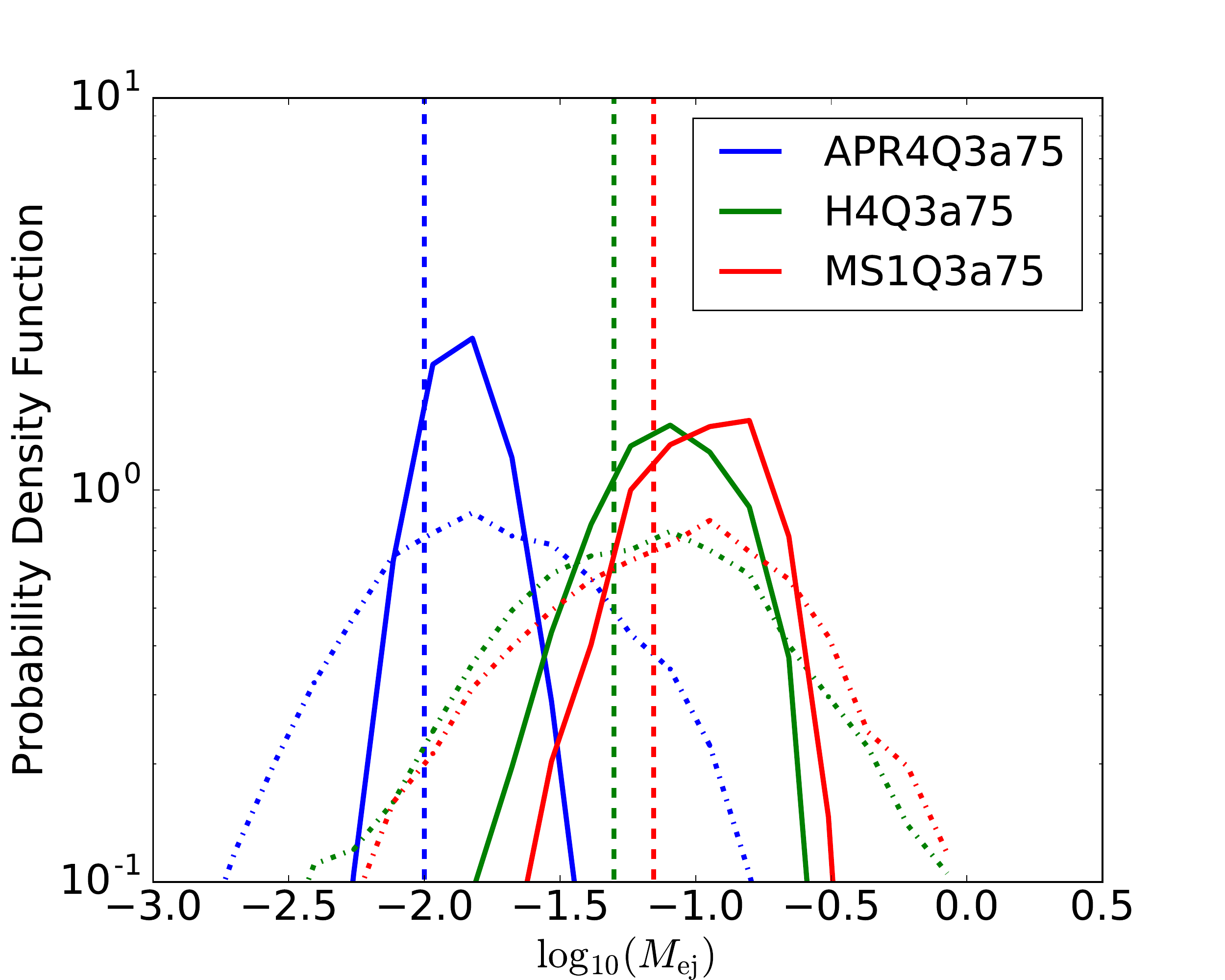} 
  \caption{
   Histograms of $M_{\rm ej}$ for the BNS and BNHS lightcurves from 
   \cite{TaHo2014} compared with the \cite{DiUj2017} model (left) and
   \cite{KaKy2016} model (right), respectively.
   For this analysis, there are errors of 0.2\,mag (solid lines)
   and errors of 1\,mag (dash-dotted lines).}
 \label{fig:BNS_BHNS_Tanaka_mej}
\end{figure*}

We now perform a comparison between the parameterized models and results 
from \cite{TaHo2014}. For this analysis, we distinguish between BNS and 
BHNS. The BNS setups of \cite{TaHo2014} are compared to the 
\cite{DiUj2017} model and the BHNS lightcurves are compared to the 
\cite{KaKy2016} model. In Figure~\ref{fig:BNS_BHNS_Tanaka_mej}, we show 
histograms for $M_{\rm ej}$ for uncertainties of 1\,mag (dash-dotted 
lines), which corresponds to the error stated in \cite{KaKy2016} and 
\cite{DiUj2017}. The ejecta mass corresponding to the lightcurves of 
\cite{TaHo2014} (vertical dashed lines) is always within the posteriors 
of the models for the 1\,mag posteriors (dash-dotted). We find that for 
0.2\,mag (solid lines) uncertainties some of the true values for BNS 
systems lie outside the estimated posteriors, which is to be expected 
because the uncertainties in \cite{DiUj2017} and \cite{KaKy2016} are 
1\,mag. But, even for an assigned uncertainty of 0.2\,mag, the 
posteriors of the BHNS setups are consistent with the injected values, 
which suggests that recovering smaller ejecta masses is in general less 
accurate. This might be caused by inaccuracies in the employed 
bolometric corrections and is already visible in Figure~9 of 
\cite{DiUj2017}.

\subsection{Comparison with other kilonova models}

\begin{figure}[t]
\centering
 \includegraphics[width=3.5in]{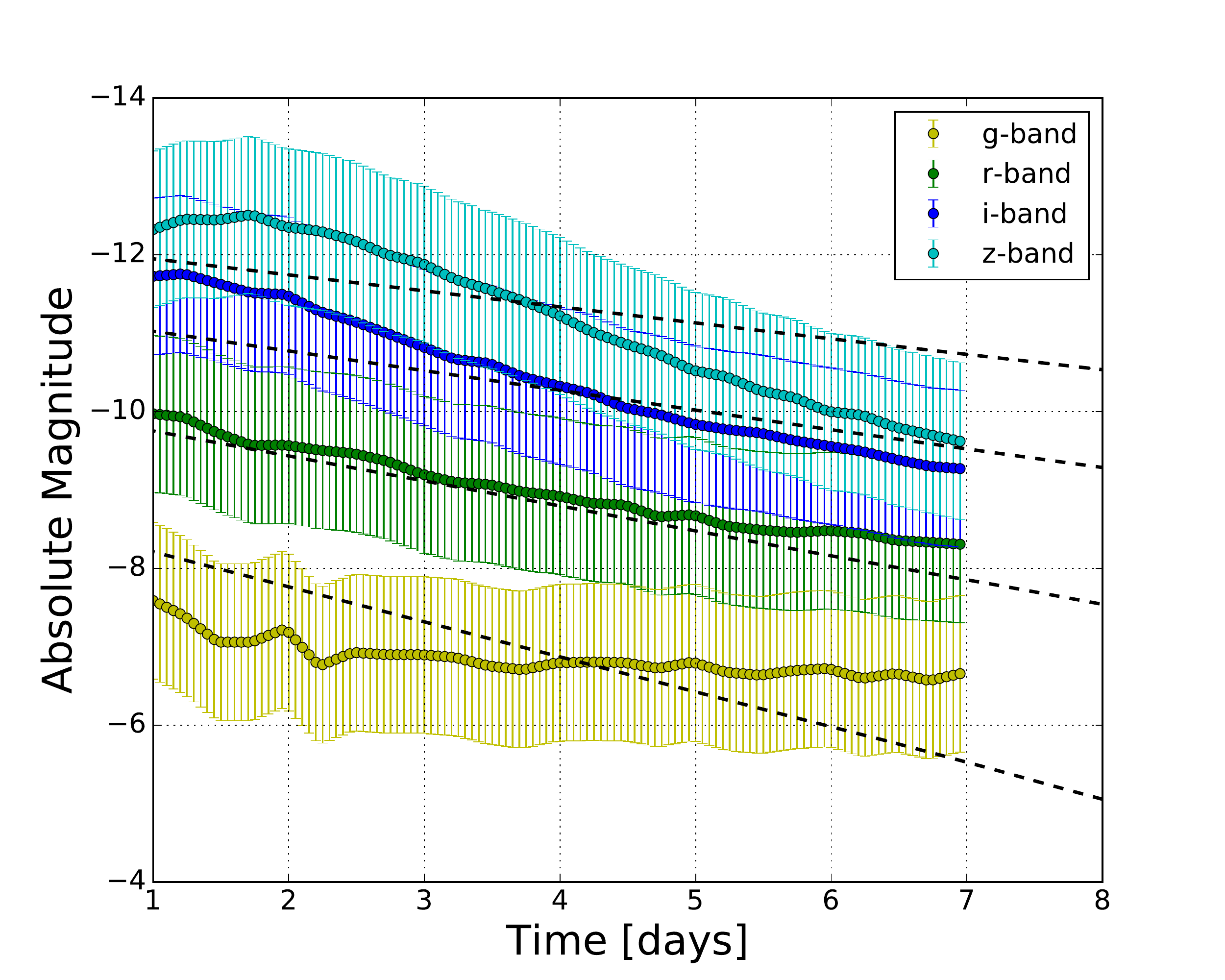} 
  \includegraphics[width=3.5in]{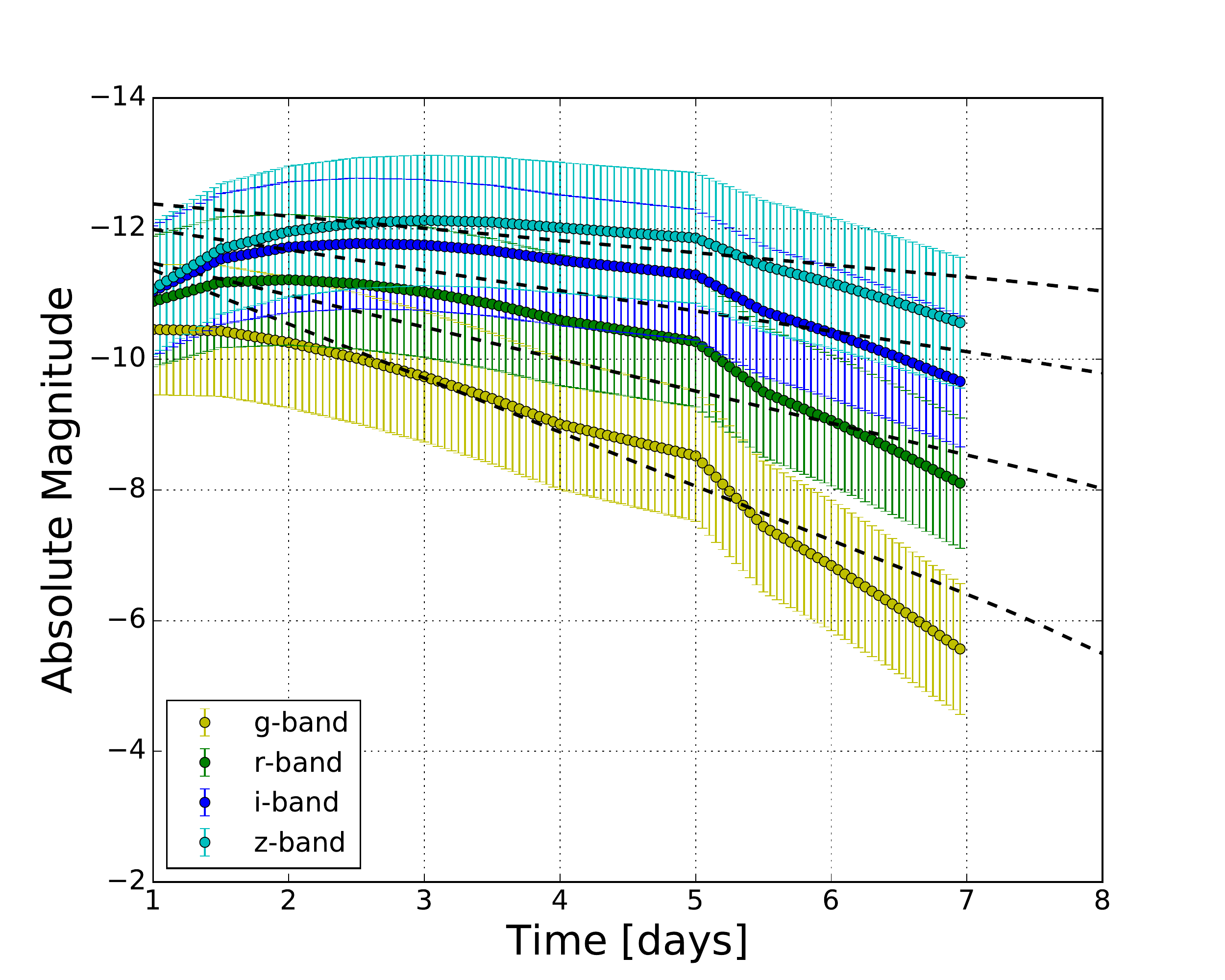}
  \caption{
   Lightcurves for \cite{BaKa2016} (top) and \cite{RoFe2017} (bottom) with the 
   same parameters as from Figure~\ref{fig:models}. 
   We also perform a maximum likelihood chi-squared fit to each 
   lightcurve using the \cite{DiUj2017} model for comparison.
   }
 \label{fig:BNS_Barnes_Rosswog}
\end{figure}

We now perform a comparison between the parametric models and 
\cite{BaKa2016,RoFe2017}. In Figure~\ref{fig:BNS_Barnes_Rosswog}, we 
take the \cite{BaKa2016} (top panel) model rpft\_m005\_v2 and the NS14B7 
model of \cite{RoFe2017} (bottom panel), and use the \cite{DiUj2017} 
model for recovery. One finds that the relative magnitudes between the 
bands is mostly consistent across the models. However, the models are 
not able to reproduce the lightcurves as accurately as for 
\cite{TaHo2014}. We find that multiple parameters, including the ejecta 
mass, cannot be constrained. Furthermore, for \cite{RoFe2017} the 
parameter estimation pipeline leads to a $T_0$ estimate of the order of 
a few days, which suggests that follow up searches using the current 
parametrized models would not correctly detect transients with 
lightcurves similar to those given in \cite{RoFe2017}.

The origin of the difference between the parameterized models and 
\cite{BaKa2016,RoFe2017} is that \cite{DiUj2017} was built using the 
lightcurves of \cite{TaHo2014}. It can be expected that parametrized 
models approximating the results of \cite{BaKa2016} and \cite{RoFe2017} 
can be obtained as well. This shows that for future development, it is 
urgently required to provide lightcurves using full radiative transfer 
simulations that are as realistic as possible, i.e.~including different 
ejecta components, time dependent efficiency, and complex ejecta 
morphologies.

\subsection{Comparison with other models}

\begin{figure}[t]
\centering
 \includegraphics[width=3.5in]{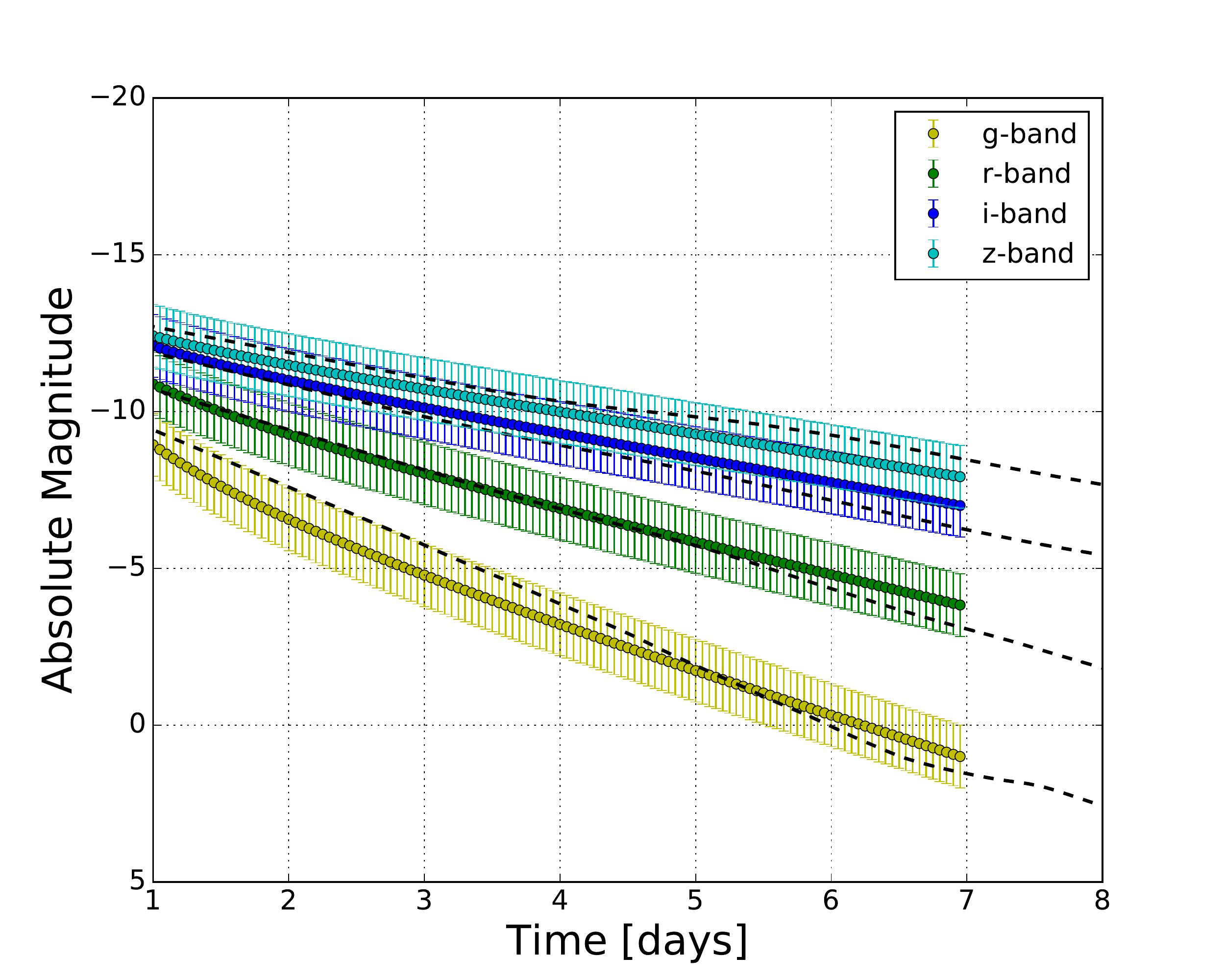}
 \includegraphics[width=3.5in]{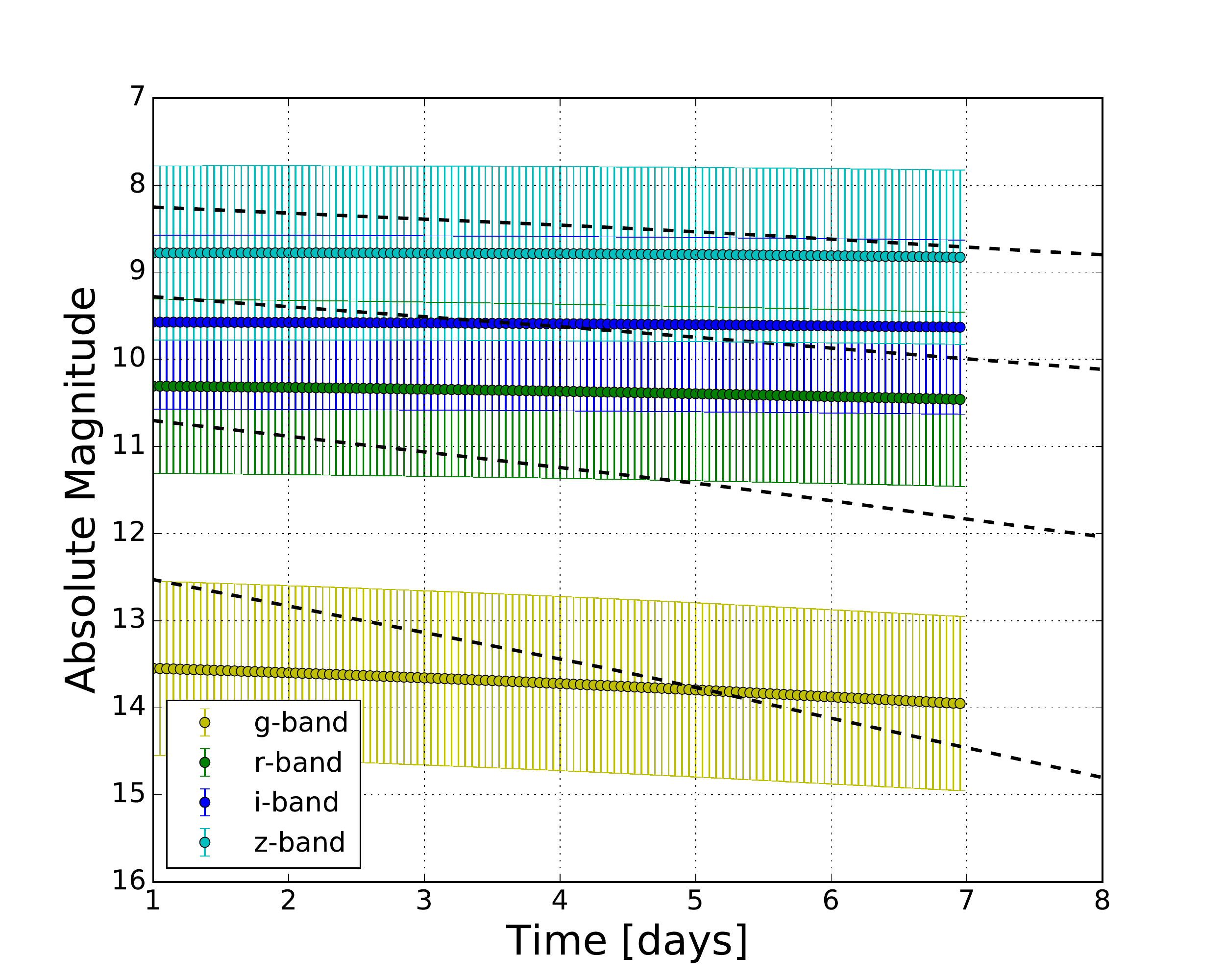} 
  \caption{
   Lightcurves for \cite{MeBa2015} (top), with the same parameters as from 
   Figure~\ref{fig:models}, and a SN Ia from \cite{GuAs2007} (bottom). 
   We also perform a maximum likelihood chi-squared fit to each 
   lightcurve using the \cite{DiUj2017} model for comparison.}
 \label{fig:BNS_Metzger_SN}
\end{figure}

We also compare to a few non-kilonova models in Figure~\ref{fig:BNS_Metzger_SN}.
Considering the different origin of the EM signal, 
we expect that the kilonova models cannot capture the injected lightcurves. 
We use the \cite{MeBa2015} fiducial model (top panel of Figure~\ref{fig:BNS_Metzger_SN}) 
describing the blue kilonovae precursor and a SN Ia model from \cite{GuAs2007} (bottom panel). 
\cite{MeBa2015} does have an initially higher blue component.
The best fit curve from \cite{DiUj2017} is capable of producing time 
dependent lightcurve approximants.
For the SN Ia it was not possible to compute time dependent lightcurves with the 
parametrized models which approximate the SN Ia lightcurve.
This shows that the parametrized models can also help to distinguish
transients with different origins. 

\section{Extracting the binary parameters} \label{extracting}

Our previous study focused on the question how we can use parametrized models 
to obtain information about the mass, velocity, and morphology of the ejecta. 
At least as important for astrophysical considerations is 
the question whether measured lightcurves can be used to directly constrain
the binary properties: masses, spins, and possibly also the unknown EOS.  
To achieve this goal, phenomenological models connecting the ejecta properties 
as well as the binary parameters have to be employed. Such models based on large sets of 
NR simulations are given in \cite{KaKy2016} for BHNS systems and in \cite{DiUj2017} for BNS systems.
Because of the large uncertainties in the determination of the ejecta mass in full general relativistic 
simulations, current parametrized models can only be seen as a starting point to more accurate models.
Longer simulations with detailed microphysics 
are needed to properly model all the ejecta components. 

\subsection{Possible Degeneracies}

\begin{figure}[t]
\centering
 \includegraphics[width=3.0in]{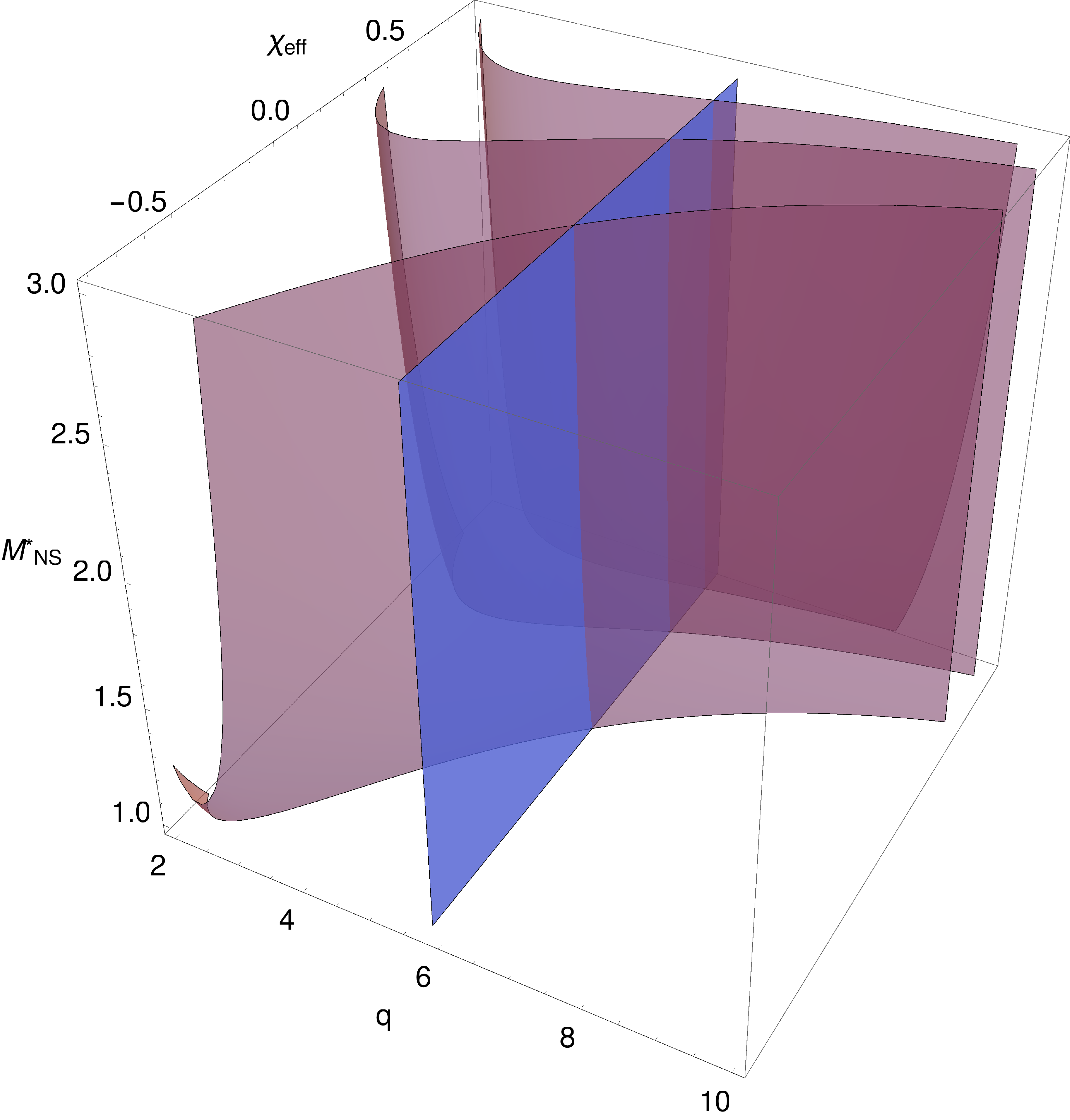}
  \caption{
  Binary parameters of a BHNS systems, Equation~\ref{eq:BHNS:mej_fit}
  and Equation~\ref{eq:BHNS_vfit}, which lead to $M_{\rm ej} = 10^{-2}$ 
  (red surfaces) and $v_{\rm ej}=0.28$ (blue surface) under the assumption 
  of different compactnesses $C=0.13,0.15,0.17$ (from left to right). Because of 
  the degeneracies between the binary parameters and the ejecta properties an unambiguous
  measurement of $q,M_{\rm NS}^*,\chi_{\rm eff},C$ is not possible if only $M_{\rm ej}$ and $v_{\rm ej}$
  are measured.  
 }
 \label{fig:degeneracies}
\end{figure}

In addition to the large uncertainty of the NR data, 
the models also contain degeneracies which do not 
allow the simultaneous extraction of all 
binary parameters. 
The ejecta mass and velocity as functions of the binary 
parameters for BHNS can be approximated by: 
\begin{small}
\begin{eqnarray}
	\frac{M_{\rm ej}}{M_{\rm NS,*}}& = 
	&{\rm Max}\left\{a_1 q^{n_1}\frac{1-2 C}{C}-a_2\,q^{n_2}\,{\tilde r}_{\rm ISCO} (\chi_{\rm eff}) \right. \label{eq:BHNS:mej_fit}  \\
	 &+& \left. a_3\left(1-\frac{M_{\rm NS}}{M^*_{\rm NS}}\right)+a_4,0 \right\},  \nonumber \\
	v_{\rm ej}&=& b_1 \,q+b_2 \label{eq:BHNS_vfit}.
\end{eqnarray}
\end{small}

\noindent with $\chi_{\rm eff}=\chi\,{\rm cos}\,i_{\rm tilt}$, where 
$i_{\rm tilt}$ 
is the angle between the dimensionless spin of the black hole $\chi$ and the orbital angular momentum and 
${\tilde r}_{\rm ISCO}$ is the radius of the innermost stable circular orbit normalized by the black hole mass. 
$a_1,a_2,a_3,a_4,n_1,n_2,b_1,b_2$ are fitting parameters which are determined by comparison 
to a large set of NR data, see \cite{KaKy2016}.
For BNS setups, the ejecta properties are approximated by 
\begin{small}
\begin{eqnarray}
 M_{\rm ej}^{\rm fit} & = & 10^{-3} \left[
 \left\{ a \left(\frac{M_2}{M_1}\right)^{1/3} \left(\frac{1-2 C_1}{C_1}\right)+ 
 b \left(\frac{M_2}{M_1}\right)^{n} \right. \right. \label{eq:BNS:Mej_fit} \\ 
 &+& \left. \left. c \left(1 - \frac{M_1}{M^*_{1}}\right) \right\} M_1^* + 
 (1\leftrightarrow 2) + d \right],  \nonumber \\
v_{\rm ej} & = & \sqrt{v_\rho^2+v_z^2},  \label{eq:BNS:vej} \\
 v_{\rho,z} & = & \left[ a_{\rho,z} \left(\frac{M_1}{M_2}\right) \left(1 + c_{\rho,z}\  C_1\right)  \right] 
 +  (1\leftrightarrow 2) + b_{\rho,z}, \label{eq:BNS:vrho_fit}  \\
 \theta_{\rm ej} & = & \frac{ 2^{4/3} v_\rho^2 - 2^{2/3} 
(v_\rho^2 ( 3 v_z + \sqrt{ 9 v_z^2 + 4 v_\rho^2}))^{2/3} }
{(v_\rho^5(3 v_z + \sqrt{9 v_z^2 + 4v_\rho^2}))^{1/3}},  \quad \label{eq:BNS:theta_fit} \\
 \phi_{\rm ej}  & = &   4 \theta_{\rm ej} + \frac{\pi}{2},  \label{eq:BNS:phi_fit}
\end{eqnarray}
\end{small}

\noindent with the fitting parameters 
$a,b,c,d,a_\rho,a_z,b_\rho,b_z,c_\rho,c_z,n$ given in \cite{DiUj2017}.

As can be concluded from the Equations~\ref{eq:BHNS:mej_fit}-\ref{eq:BNS:phi_fit}, 
the BHNS model depends on: the mass ratio $q$, the ``effective'' spin of the black hole $\chi_{\rm eff}$, 
the baryonic mass of the neutron star $M_{\rm NS}^*$, 
the quotient of the neutron star's gravitational mass $M_{\rm NS}$ 
and baryonic mass $M^*_{\rm NS}$, 
and its compactness $C$, i.e., five parameters. 
For the case of BNS systems, the number increases to six: 
the gravitational masses $M_1,M_2$, the baryonic masses 
$M_1^*, M_2^*$, and the compactnesses $C_1,C_2$ of the neutron stars.

As an example to visualize possible degeneracies in 
Equations~\ref{eq:BHNS:mej_fit}-\ref{eq:BNS:phi_fit}, let us suppose that 
the ejecta mass and the ejecta velocity was measured for a BHNS setup. 
In Figure~\ref{fig:degeneracies}, we show as red surfaces the allowed binary 
parameters for which $M_{\rm ej}=10^{-2}$ under the assumptions 
of $C=0.13,0.15,0.17$. In addition, we make use of the quasi-universal 
relation Equation~\ref{eq:quasi_univ_fit} to connect the gravitational and baryonic mass to the 
compactness, see discussion in the next subsection. As a blue surface,
we mark the binary parameters for which $v_{\rm ej} = 0.28$. According to 
Equation~\ref{eq:BHNS_vfit}, the measurement of $v_{\rm ej}$ would determine 
the mass ratio of the system $q$ but leave the other parameters unconstrained. 

Figure~\ref{fig:degeneracies} shows that even if $M_{\rm ej}$ and $v_{\rm ej}$ are
accurately known, the binary parameters cannot be determined.
The intersections between the red and blue surfaces mark all the 
allowed regions for which the ejecta properties 
are consistent with the estimated $M_{\rm ej},v_{\rm ej}$ under the assumption of 
a given compactness $C$. Consequently, an accurate measurement of the binary properties 
is only possible for cases 
for which more parameters than $M_{\rm ej},v_{\rm ej}$ are determined, 
e.g.~$\theta_{\rm ej}$ and $\phi_{\rm ej}$, or for cases where due to 
a simultaneous detection of GWs some binary parameters are known.  

\subsection{Quasi-universal properties}

\begin{figure}[t]
\centering
 \includegraphics[width=3.5in]{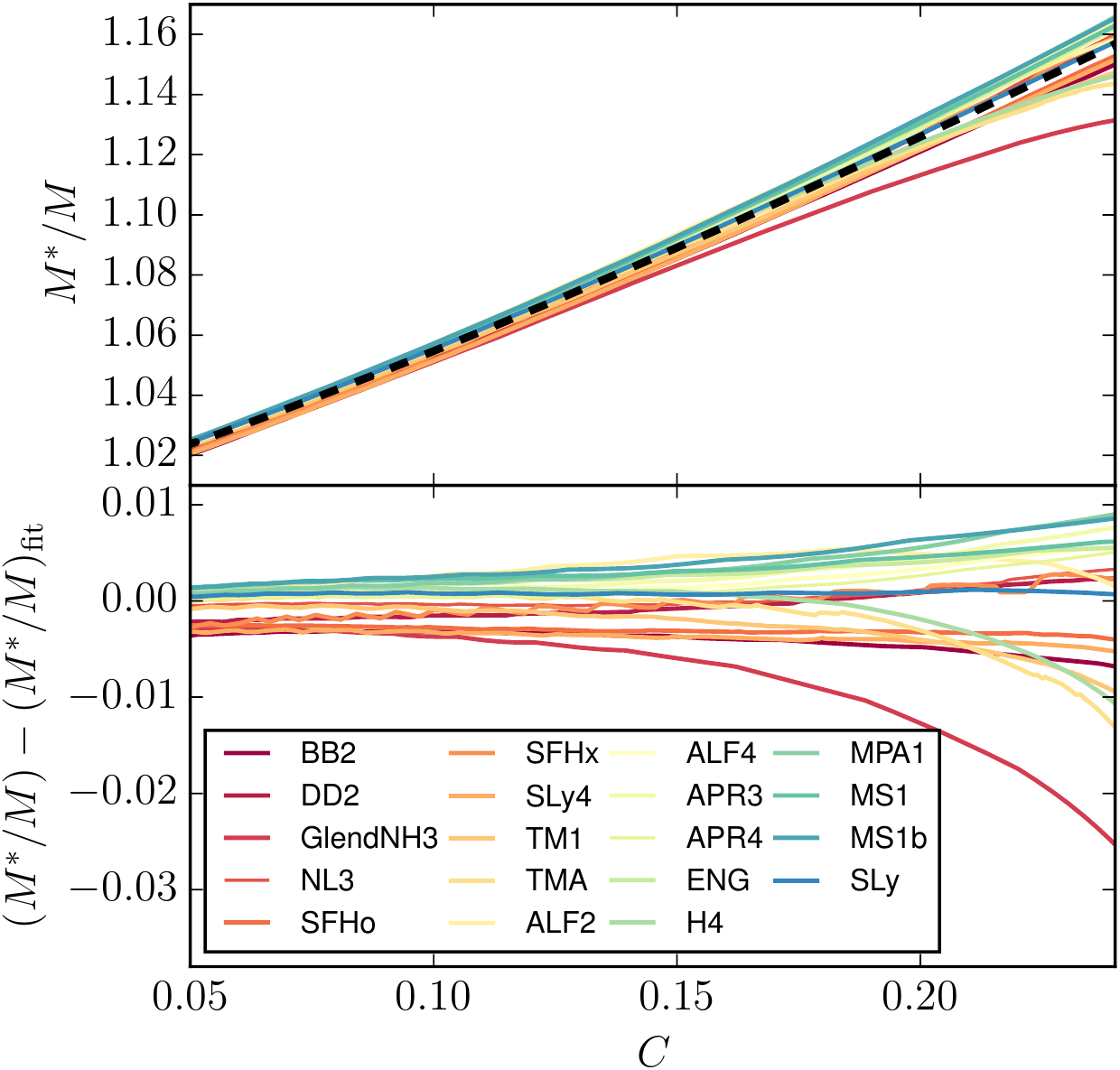}
  \caption{Ratio of the baryonic mass and gravitational mass $M^*/M$ 
  as a function of the compactness for different equations of state (top panel) 
  and difference for $M^*/M$ between each EOS and the approximant 
  Equation~\eqref{eq:quasi_univ_fit} (bottom panel). The dashed black line corresponds to the fit. 
 }
 \label{fig:quasiuniv_fit}
\end{figure}

Due to the large number of unknown binary parameters in 
Equations~\ref{eq:BHNS:mej_fit}-\ref{eq:BNS:phi_fit}, several degeneracies 
exist and binary parameters cannot be constrained uniquely. To reduce 
this effect, we substitute some parameters with the help of 
quasi-universal relations. Quasi-universal properties for single neutron 
stars have been first found by~\cite{YaYu2013} and were consequently 
studied for a variety of parameters, see 
e.g.~\cite{MaCa2013,PaAp2013,YagSt2014}; even in BNS systems 
quasi-universal relations are present, e.g.~\cite{BeNa2014}. We propose 
a relation between the quotient of baryonic and gravitational 
mass $M^*/M$ and the compactness $C$ of a single neutron star. To 
construct this relation, we use the EOSs employed for the dataset 
studied in \cite{DiUj2017}, but only consider EOSs which allow 
non-rotating NS masses above $1.9$, which lies even below the 
highest measured NS mass of $\approx 2.01$. 
Figure~\ref{fig:quasiuniv_fit} shows $M^*/M$ as a function of the 
compactness $C$ for all EOSs. We find that only a small spread is caused 
by the EOSs. Except for GlendNH3, all curves stay close together.

We fit all data with an approximant of the form
\begin{equation}
 \frac{M^*}{M}= 1 + a~C^n, \label{eq:quasi_univ_fit}
\end{equation}
the free fitting parameters are $a=0.8858$ and $n=1.2082$. The fit is 
included as a black dashed line in the top panel of 
Figure~\ref{fig:quasiuniv_fit}. By construction, we obtain for 
$C\rightarrow0$ the correct limit of $M^*/M \rightarrow 1$. The 
residuals of the fit is shown in the bottom panel of 
Figure~\ref{fig:quasiuniv_fit}. Absolute errors within the compactness 
interval of $C \in [0.05,0.24]$ are within $\pm 0.01$, except for 
GlendNH3. This leads to fractional errors of $\lesssim 10\%$ for the 
term $1-M^*/M$ which enters directly in the ejecta mass computation for 
BHNS and BNS systems. On average fractional errors are $\lesssim 3\%$. 
Considering the large uncertainty of 
Equations~\ref{eq:BHNS:mej_fit}-\ref{eq:BNS:phi_fit}, we expect that the 
error caused by Equation~\eqref{eq:quasi_univ_fit} is negligible. But by 
introducing this relation, the number of free parameters for the BHNS 
model is reduced by one and for the BNS model reduced by two. This 
allows for significantly better extraction of the binary parameters from 
the ejecta properties.

\subsection{Extraction of binary parameters}

\begin{figure*}[t]
 \includegraphics[width=3.5in]{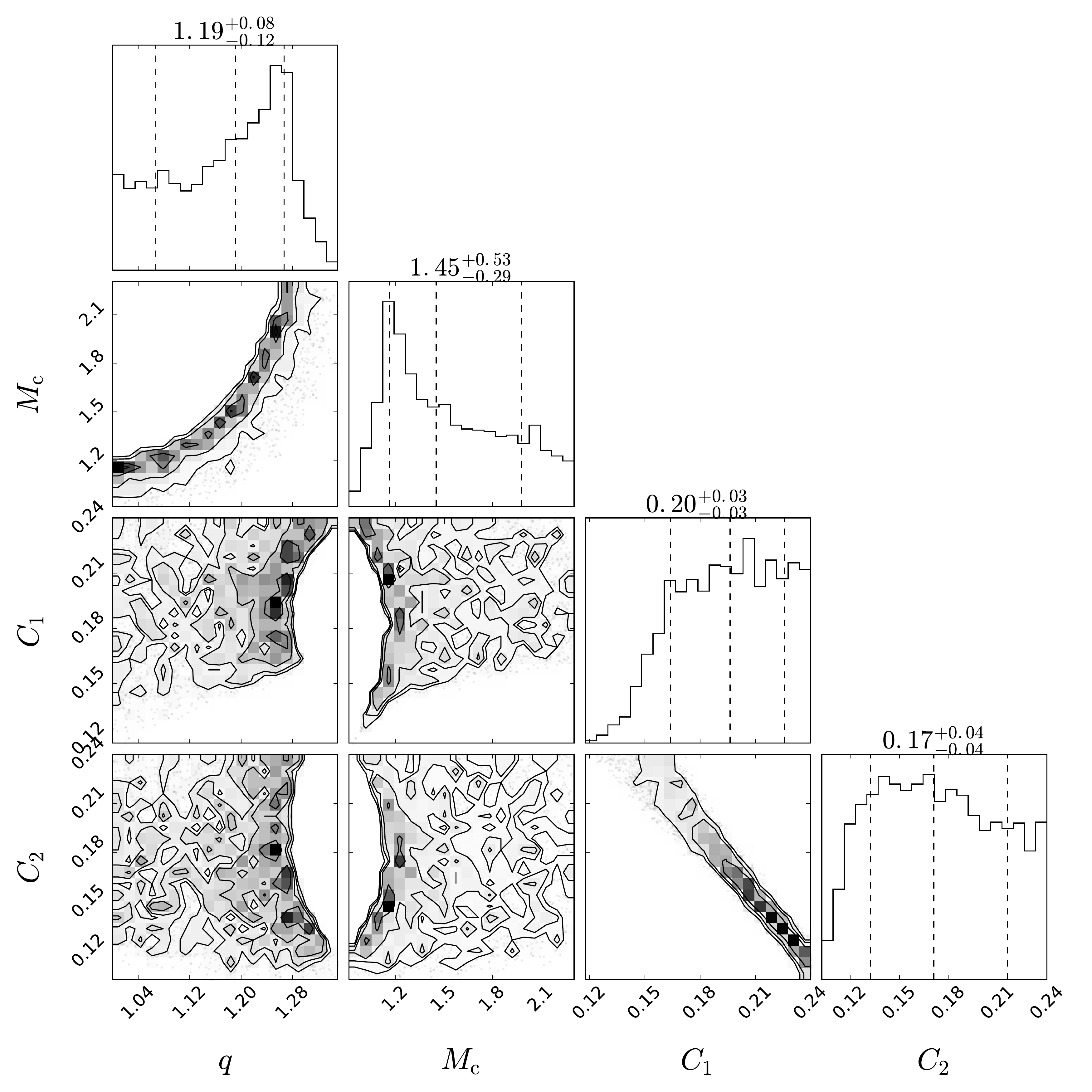}
  \includegraphics[width=3.5in]{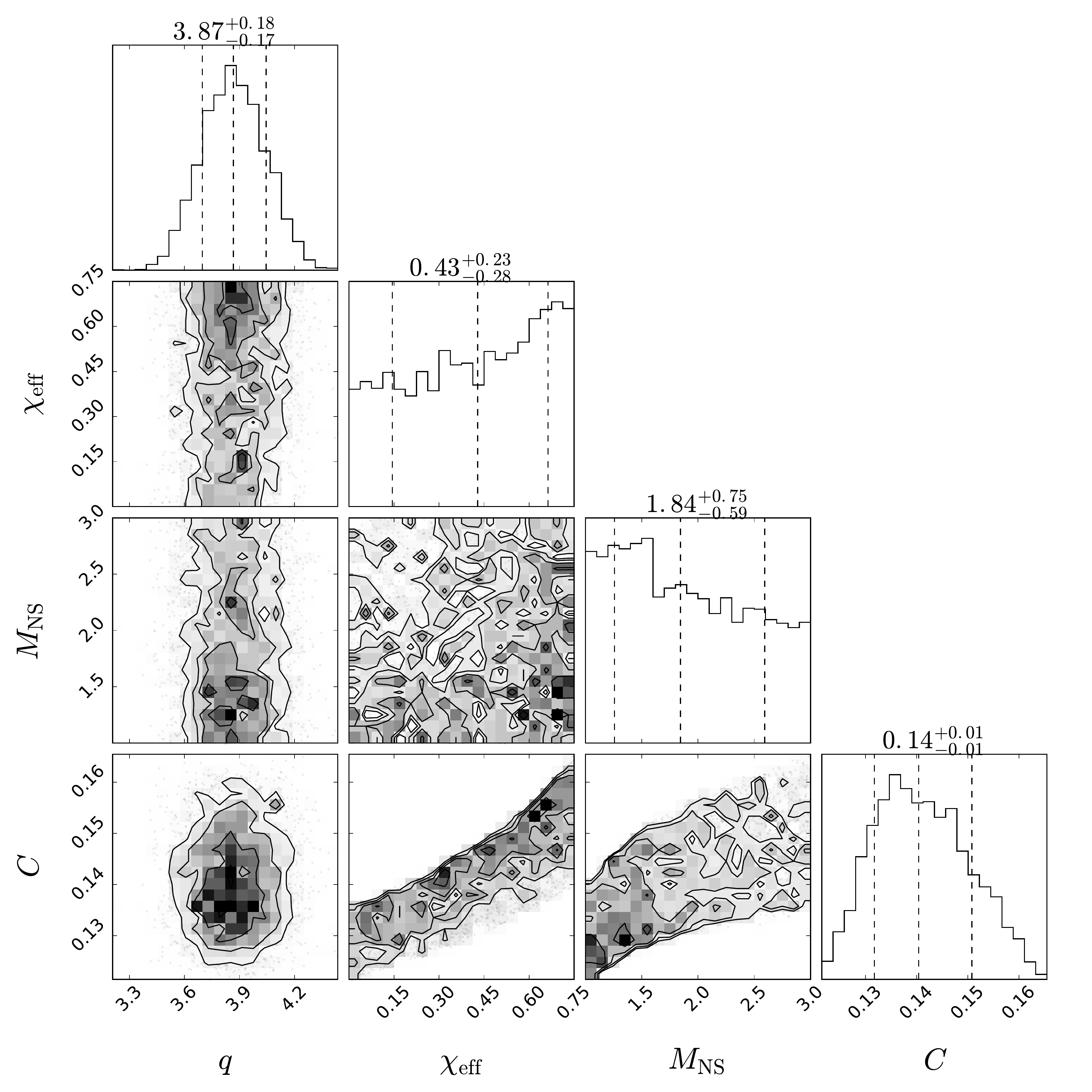} 
  \caption{
  On the left is the corner plot for the model fits 
  for the \cite{DiUj2017} model with 
  $M_{\rm ej} = 5 \times 10^{-3}$, $v_{\rm ej} = 0.25$ 
  and an optimistic 1\% Gaussian errorbar on the measurement.
  On the right is the same for the \cite{KaKy2016} model with 
  $M_{\rm ej} = 5 \times 10^{-2}$ and $v_{\rm ej} = 0.25$ for 
  comparison with the same error bars.
 }
 \label{fig:fitting}
\end{figure*}

In the following, we use a similar scheme as in Sec.~\ref{sec:PE} 
to explore how binary parameters can be recovered from a kilonovae 
detection. 
We explore the situation where we have made a measurement of $M_{\rm ej}$ and $v_{\rm ej}$.
We calculate the likelihood using a kernel density estimator commonly 
used in GW data analysis \citep{SiPr2014}.
This technique is useful for cases where the measurements of those 
distributions arise from parameter estimation with potentially 
highly correlated estimates amongst the variables, as is common in GW data analysis.
The priors used in the analyses are as follows:
For \cite{KaKy2016}, 
$3 \leq q \leq 9$, 
$0 \leq \chi_{\rm eff} \leq 0.75$, 
$1 \leq M_{\rm NS} \leq 3$, 
and $0.1 \leq C \leq 0.2$,
while for \cite{DiUj2017}, 
$1 \leq M_{1} \leq 3$, $1 \leq M_{2} \leq 3$, $0.08 \leq C_1 \leq 0.24$, 
and $0.08 \leq C_2 \leq 0.24$.
The differences in compactness prior ranges are due to the differences 
in compactness used in the simulations the models used.
The priors are flat over the stated ranges.
For this reason, significant structure in the 1D and 2D contours arise from the posterior.\\

To begin, we explore the correlation between the variables by employing 
the very optimistic assumption of 1\% Gaussian errorbars on the 
measurement, which essentially inverts the equations in the previous 
section. We show in Figure~\ref{fig:fitting} the parameters consistent 
with two different choices of $M_{\rm ej}$ and $v_{\rm ej}$. For the BNS 
case (left panel), we choose $M_{\rm ej} = 5 \times 10^{-3}$ and $v_{\rm 
ej} = 0.25$, for the BHNS case (right panel), we choose $M_{\rm ej} = 5 
\times 10^{-2}$ and $v_{\rm ej} = 0.25$. In general, for the BNS 
systems, the constraints are not strong given the relatively wide 
variety of parameters that support non-zero ejecta masses and 
velocities. We choose to plot mass ratio ($q = M_1/M_2$) and chirp mass 
[$M_{\rm c} = (M_1 M_2)^{3/5} (M_1 + M_2)^{-1/5}$] instead of $M_1$ and 
$M_2$, due to the clearer peaks in this parameterization. We clearly see 
in the 2D corner plots degeneracies between $M_{\rm c}$ and $q$ as well 
as between $C_1$ and $C_2$, which are similar to those described in the 
previous subsection. These indicate the fundamental limitations of 
EM-only observations in the measurements of these quantities. For the 
BHNS systems, the main constraint is on $q$, which has some correlation 
with compactness. Due to the significant correlations between $q$, 
$\chi_{\rm eff}$, and $C$, it will be difficult to constrain those 
parameters without measurements from other quantities.\\

\begin{figure*}[t]
 \includegraphics[width=3.5in]{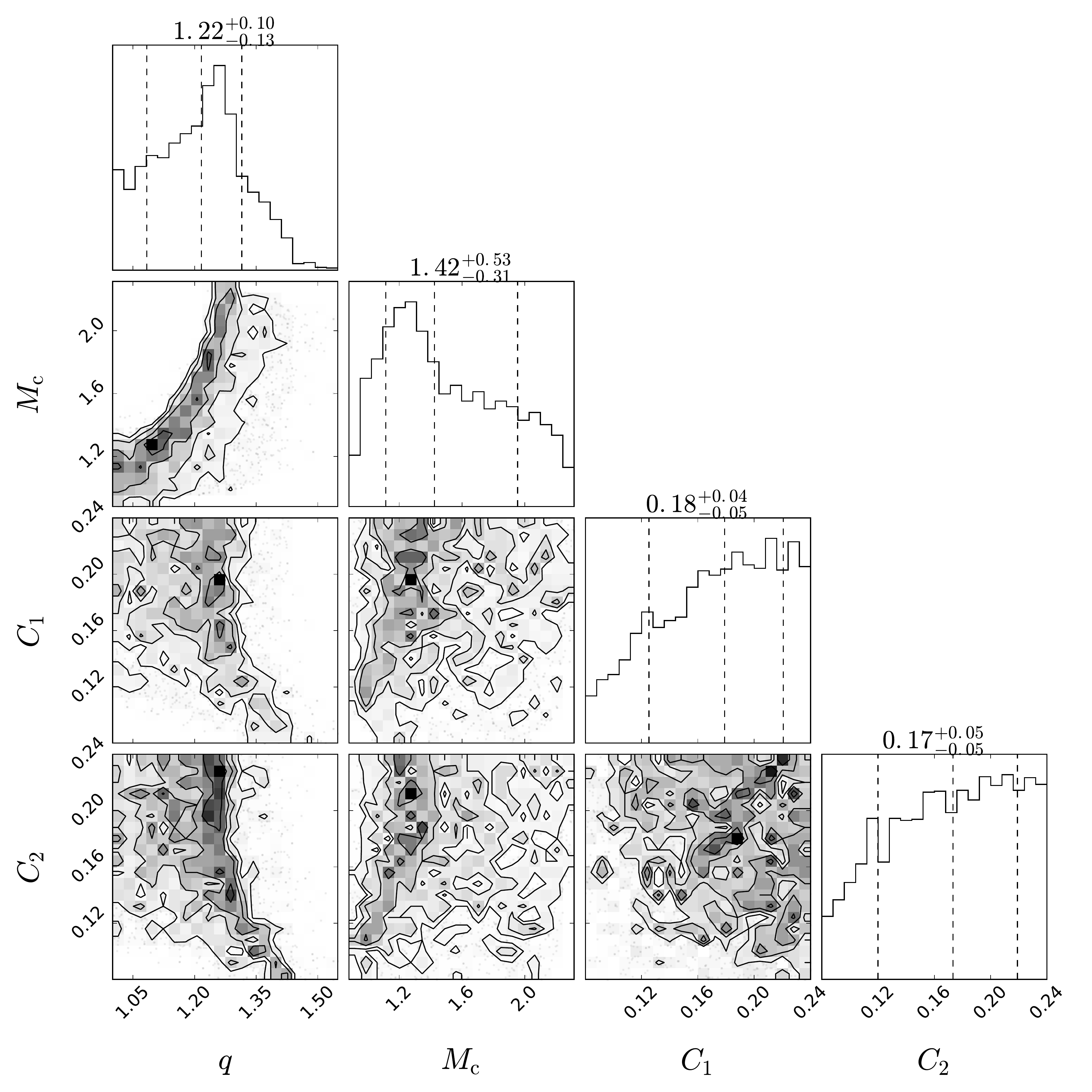}
  \includegraphics[width=3.5in]{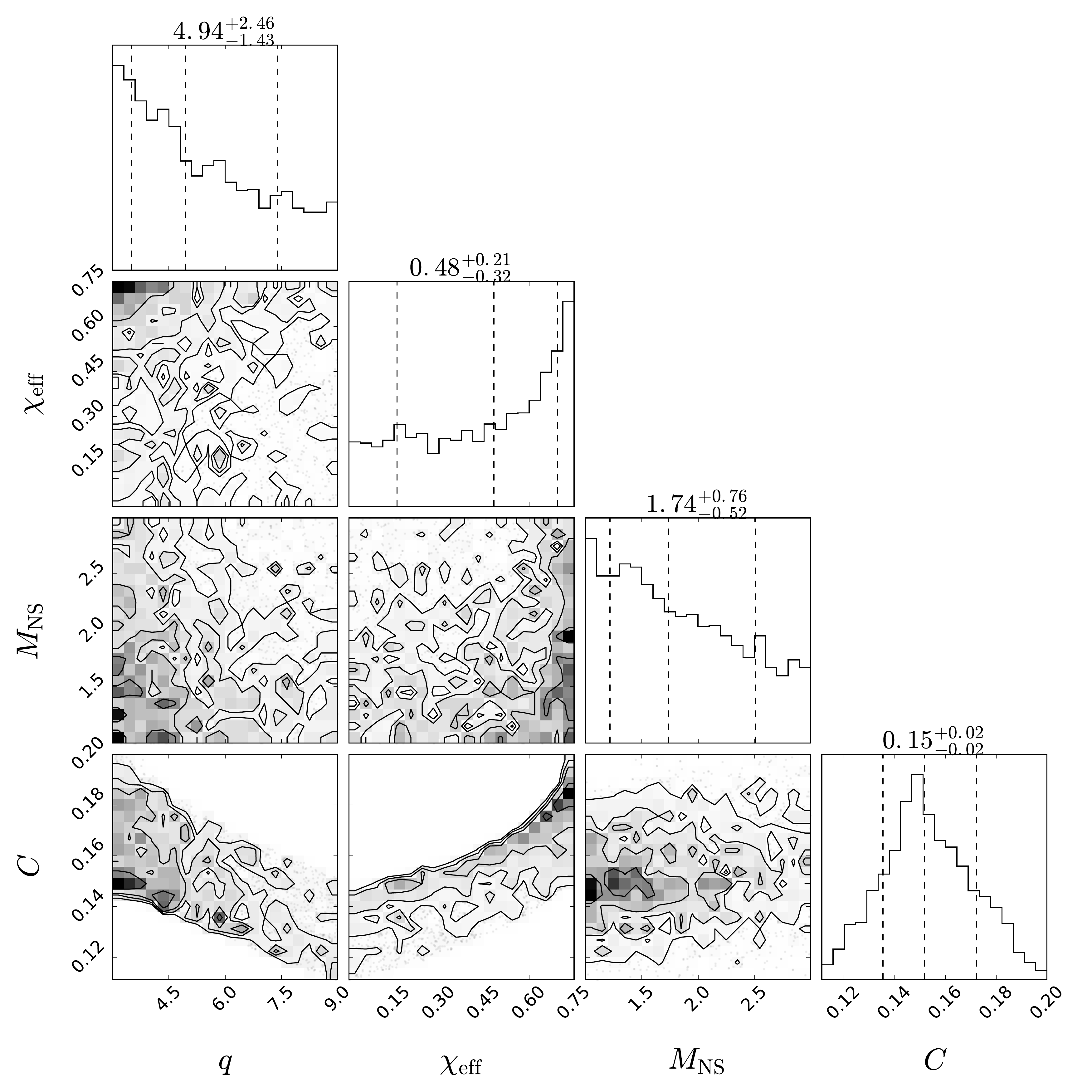} 
  \caption{
  On the left is the corner plot for the model fits for the 
\cite{DiUj2017} model for $M_{\rm ej}$ and $v_{\rm ej}$ contours sampled 
from a lightcurve with $M_{\rm ej} \approx 5 \times 10^{-3}$, $v_{\rm 
ej} \approx 0.2$ (similar to Figure \ref{fig:fitting}) and model 
uncertainties of $0.2$\,mag. On the right is the same for the 
\cite{KaKy2016} model for comparison.
 }
 \label{fig:fitting_realistic}
\end{figure*}

Figure~\ref{fig:fitting_realistic} shows more realistic levels of parameter estimates using the $M_{\rm ej}$ and $v_{\rm ej}$ contours sampled from a lightcurve with $M_{\rm ej} \approx 5 \times 10^{-3}$
and $v_{\rm ej} \approx 0.2$ with model uncertainties of $0.2$\,mag.
The main difference in these results and the optimistic assumptions above is the relatively poor constraints on $v_{\rm ej}$.
For the BNS system (left panel), because the constraints on mass ratio are tied to $v_{\rm ej}$, most values of mass ratio are allowed in this particular case.
There are only minimal constraints on $M_{\rm c}$, $C_1$, and $C_2$. 
For the BHNS system (right panel), the only structure visible is the correlation between $q$, $\chi_{\rm eff}$, and $C$.
In case of precise measurement of the mass ratio and effective spin by 
GW parameter estimation, constraints on the neutron star 
compactness of $C \pm 0.2$ is possible.\\

\begin{figure}[h!]
 \includegraphics[width=3.5in]{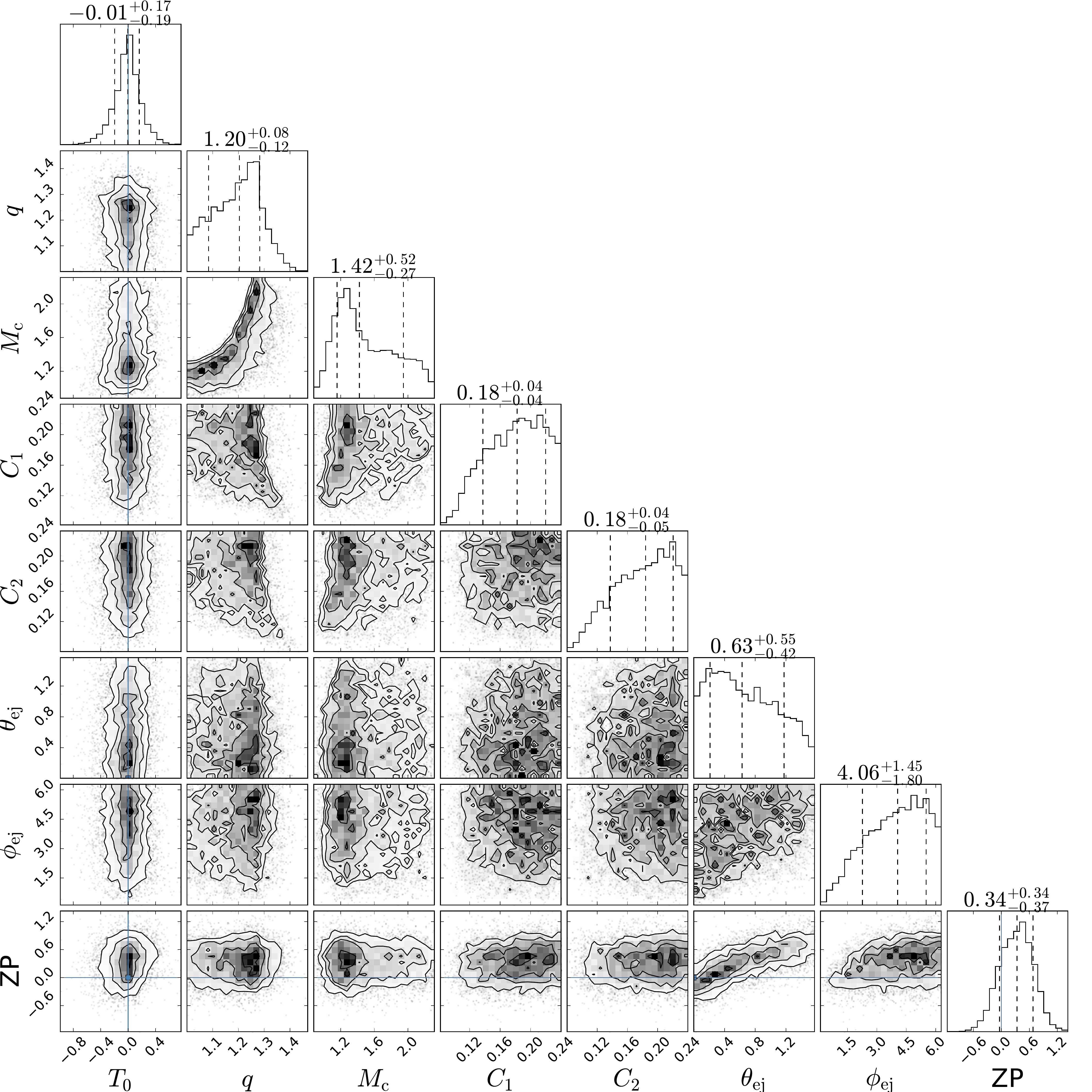}
  \caption{
   Corner plots for lightcurves with $M_{\rm ej} = 5 \times 10^{-3}$ 
   $v_{\rm ej} = 0.2$, $\theta_{\rm ej} = 0.2$\,rad, and 
   $\phi_{\rm ej} = 3.14$\,rad using the \cite{DiUj2017} model with 0.2\,mag 
   uncertainty.
 }
 \label{fig:BNS_masses_fit}
\end{figure}

As a final comparison, we perform parameter estimation for the \cite{DiUj2017} model with 0.2\,mag uncertainty, but 
instead of sampling in $M_{\rm ej}$ and $v_{\rm ej}$, 
we sample directly in the system parameters making use 
of Equations~\ref{eq:BNS:Mej_fit}-\ref{eq:BNS:phi_fit}.
Figure~\ref{fig:BNS_masses_fit} shows the corner plots for this scenario.
We find that the individual binary parameters are almost undetermined, only 
in the 2D $M_1$-$M_2$ or, as shown in the figure, 
$M_{\rm c}$-$q$ plane a clear contour is visible.
According to the 1D posteriors of $q$ it seems that high 
mass ratios are ruled out. 
Additionally, $C_1,C_2$ are almost unconstrained, but 
there seems to be a small preference for larger 
compactnesses for the shown example.  \\

Although we have only discussed the extraction of binary parameters for BNS configurations 
similar results are obtained for BHNS systems.

\section{Synergy of electromagnetic and gravitational wave observations}
\label{sec:GWEM}

\begin{figure*}[t]
 \includegraphics[width=3.5in]{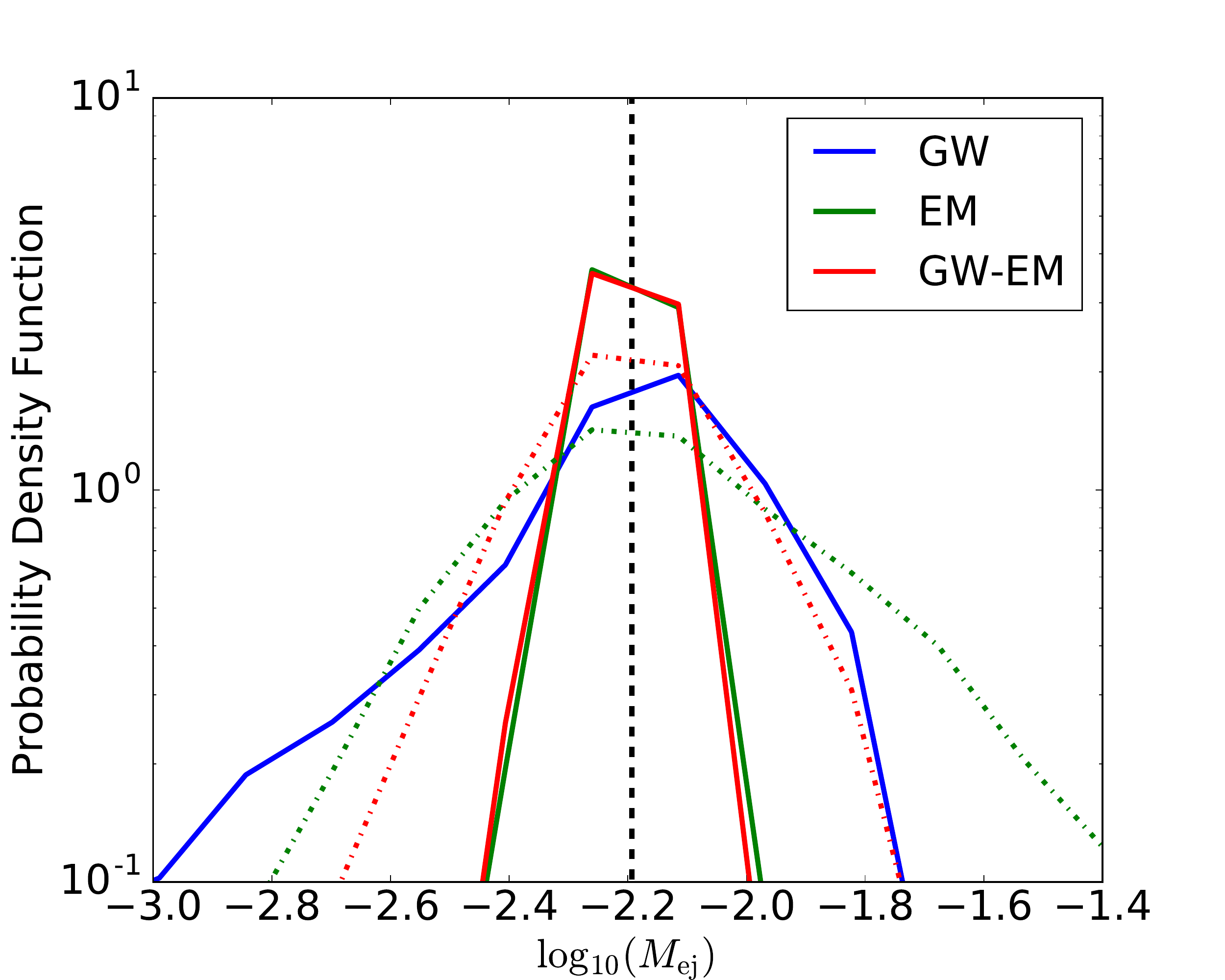}
  \includegraphics[width=3.5in]{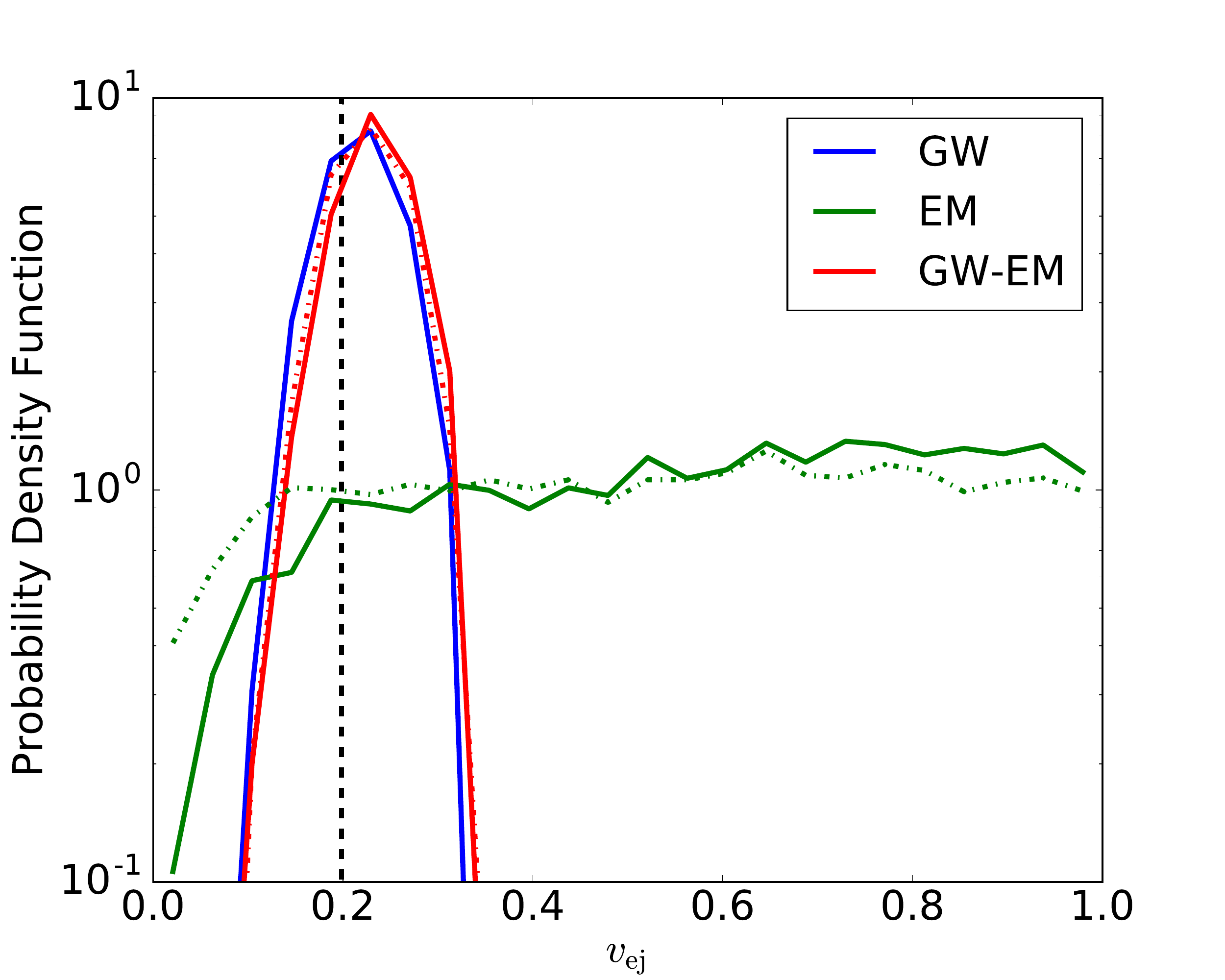} 
 \includegraphics[width=3.5in]{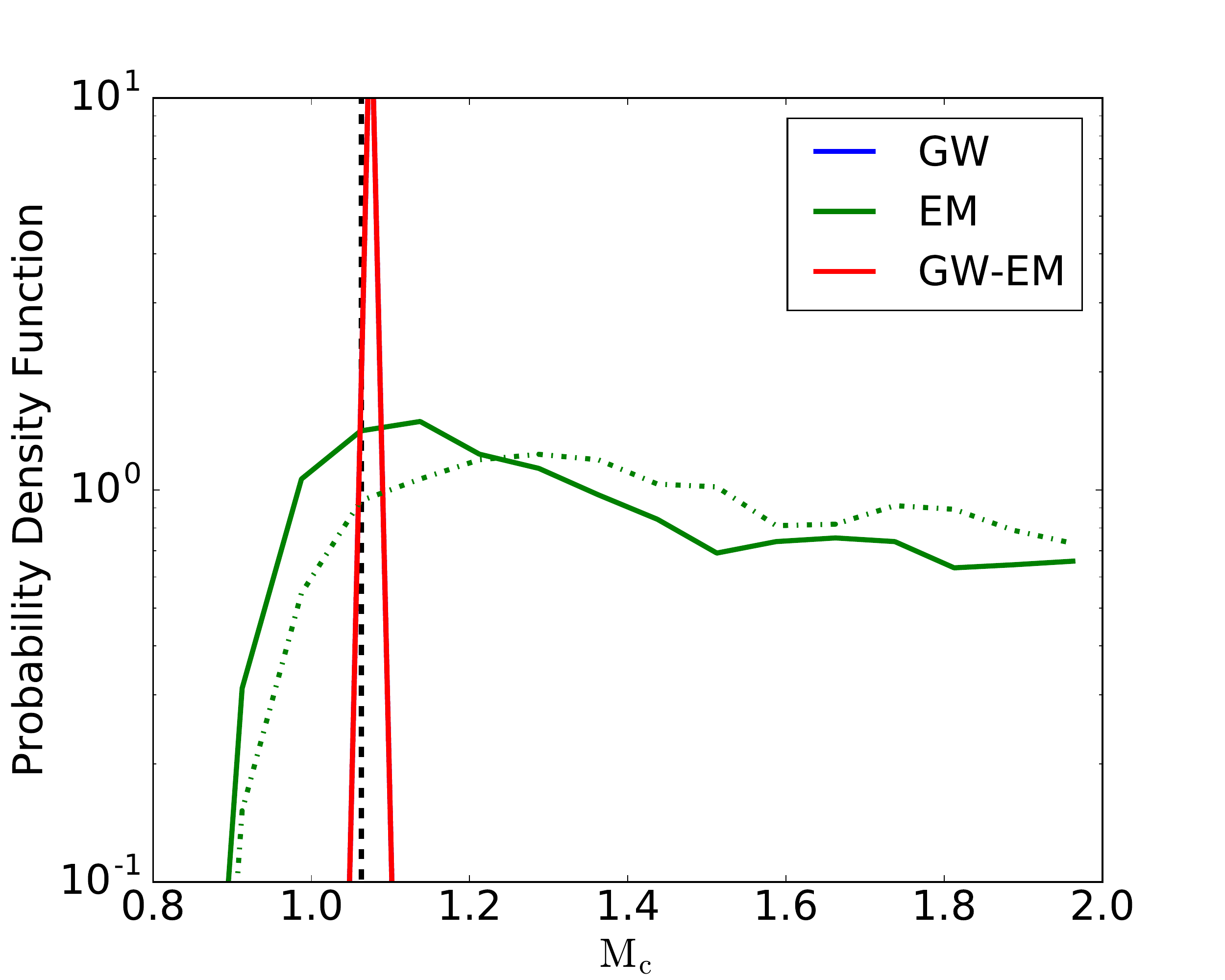}
  \includegraphics[width=3.5in]{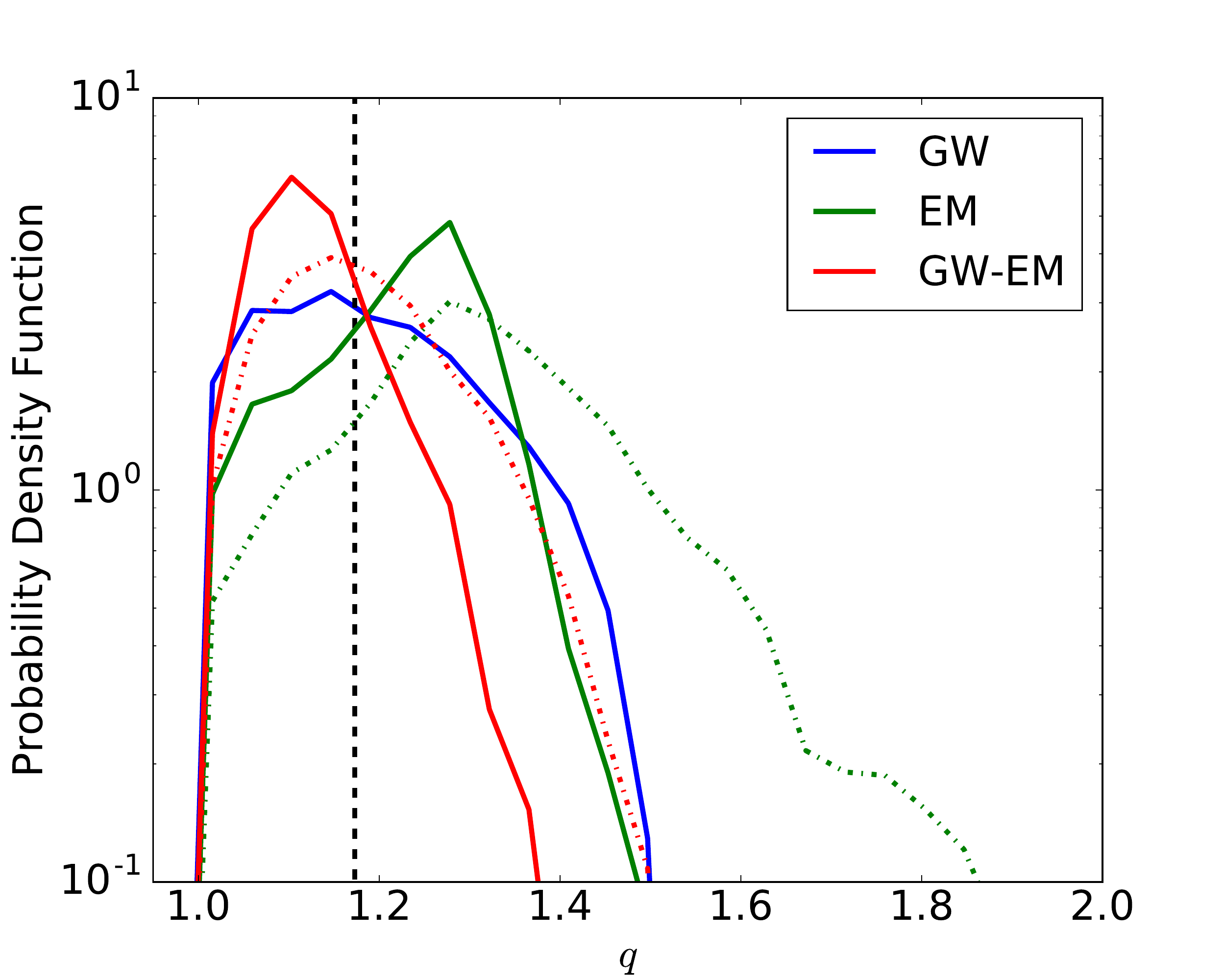}   
  \caption{
     Histograms of $M_{\rm ej}$ (top left), $v_{\rm ej}$ (top right), $M_{\rm c}$ (bottom left), and $q$ (bottom right) for EM-only, 
     GW-only and combined EM-GW constraints on a simulated BNS with GW parameter estimation from \cite{SiPr2014}.
     Parameter estimation using a simulated lightcurve from the \cite{DiUj2017} model consistent with this simulated BNS was used to generate the EM constraints. For this analysis we assume 0.2\,mag (solid lines)
   and 1\,mag (dash-dotted lines) uncertainties of the kilonova model. 
   The injected (true) value is marked as a vertical dashed line.
   In the case of $M_{\rm c}$, the GW-only line lies directly below the GW-EM line.
 }
 \label{fig:EM_GW}
\end{figure*}

As described in the previous section, the constraints on $M_{\rm c}$ and $q$ from EM observations alone are limited. On the other hand, GW 
parameter estimation provides direct constraints on these quantities as well. In particular, $M_{\rm c}$ is strongly constrained
\citep{AbEA2016a,AbEA2016g,AbEA2017}. Previously, the idea of using EM transients as triggers in searches for GWs from compact binary mergers was proposed 
\citep{KeMa2013}. Also, the possibility of combining host galaxy identification with GW parameter estimation to yield improved constraints on 
binary inclination have been mentioned before \citep{FaMe2014}. 
Additionally, we can use information from the GW parameter estimation combined with constraints from the 
EM parameter estimation to improve limits on the ejecta properties.

To demonstrate the benefits of this kind, we take an example from \cite{SiPr2014}, which includes both GW skymaps and posteriors from the 
parameter estimation of BNS signals. We take one such example and generate a lightcurve using the \cite{DiUj2017} model corresponding to the 
mean of the mass posteriors with compactnesses of $C_{1,2} = 0.147$, and use the quasi-universal relation, Equation~\ref{eq:quasi_univ_fit}, 
to compute the baryonic masses. The true values are $M_{\rm ej} = 0.006$ and $v_{\rm ej} = 0.2$. We use magnitude uncertainties of 
1.0\,mag and 0.2\,mag.

We perform the same parameter estimation technique as in the previous sections to derive EM-only constraints on $M_{\rm ej}$ and $v_{\rm ej}$. 
We then use the GW parameter estimation posteriors of $M_1$ and $M_2$ to derive GW-only constraints on $M_{\rm ej}$ and $v_{\rm ej}$. This is 
accomplished by using a kernel density estimator on the GW posteriors of $M_1$ and $M_2$ and allowing $C_1$ and $C_2$ to vary using the same 
priors as with the EM parameter estimation. Combining these posteriors is performed straightforwardly by multiplying the probabilities derived 
from both the GW-only and the EM-only posteriors, but note that 
because we are multiplying 2-D probabilities from correlated variables, the marginalized posteriors from the combined analysis can look different from multiplying the 1-D marginalized distributions.

In Figure~\ref{fig:EM_GW}, we show histograms for $M_{\rm ej}$, $v_{\rm ej}$, ${\rm M}_{\rm c}$, and $q$ for EM-only (green), GW-only (blue) and combined EM-GW (red) 
constraints. The figure demonstrates that significant improvements are possible with joint EM- and GW-parameter estimation. For example, 
whereas there are almost no limits on $v_{\rm ej}$ with EM-only, constraints from GW parameter estimation create a clear peak in the 
posterior and the ejecta velocity can be determined up to $v_{\rm ej}\approx \pm 0.15$. The limits on $M_{\rm ej}$ 
show the true synergy between potential EM and GW parameter estimation. The broad posteriors of the EM-only and GW-only are narrowed 
when combined, e.g., for an uncertainty of 1.0\,mag the uncertainty decreases from $\log_{10} M_{\rm ej}\approx \pm 0.75$ to $\log_{10} M_{\rm ej}\approx \pm 0.4$. 
In the case where a magnitude uncertainty of 0.2\,mag is employed, the constraints on velocity are still dominated by the GW parameter estimation, but the $M_{\rm ej}$ determination is dominated from the EM measurement.

Considering the binary parameters, we find that for 0.2\,mag and 1.0\,mag the chirp mass $M_{\rm c}$ is 
purely constrained by the GW parameter estimation.
On the other hand, while for a magnitude uncertainty of 1\,mag, the mass ratio is mostly determined by GW parameter estimation with only minor improvement once EM parameter estimation is also considered, one finds that for magnitude uncertainty of 0.2\,mag, constraints are improved and decrease from $q \approx \pm 0.25$ to $q \approx \pm 0.2$.
Due to the minimal correlation between $M_{\rm c}$, $q$ and the compactnesses, improved constraints on the compactnesses is not expected.\\
It is important to note that there is no bias in the measurement of $q$ in the GW-EM case.
The 1-D posterior for $q$ shifts left as the EM error bars are reduced due to the significant correlation between $M_{\rm c}$ and $q$ from the parameter estimation, as can be seen from the left of Figure~\ref{fig:fitting_realistic}.

In summary, $M_{\rm c}$ and $v_{\rm ej}$ can be constrained 
by GW parameter estimation, with little improvement from 
the inclusion of EM results. 
On the other hand, with the uncertainty budgets of current kilonova 
models and relations between binary parameters and ejecta properties, 
combined GW-EM parameter estimation improves possible constraints 
for both $M_{\rm ej}$ and $q$.
While it is true that in a future where kilonova models have 
improved such that their uncertainties are at the order of 
observation level, the EM observations will dominate the 
$M_{\rm ej}$ and $q$ constraints and therefore a combined 
analysis would not be useful, however, it is unlikely that such big improvements 
can be made in the near future. 
This motivates the 
importance for coordination between GW and EM parameter 
estimation in the event of a kilonova counterpart detection.

\section{Conclusion}
\label{sec:Conclusion}

In this article, we compared different lightcurve models, 
outlined differences and similarities, and checked the consistency amongst the models. 
We showed how parameter estimation based on the kilonovae lightcurves
depends on the uncertainty of the employed models.

We found that the parametrized models of \cite{KaKy2016} and 
\cite{DiUj2017} are able to recover the lightcurves and parameters of 
the radiative transfer simulations of \cite{TaHo2014}. As we have shown in 
Figures~\ref{fig:BNS_BHNS_error_mej} and~\ref{fig:BNS_BHNS_Tanaka_mej}, 
the ejecta properties can be determined accurately once the models have 
small uncertainty, e.g., an estimate of the ejecta mass of $\log_{10} 
M_{\rm ej}\approx \pm 0.5$ could be obtained once the model's 
uncertainty is below $0.2\,$mag. We find that currently both the 
\cite{KaKy2016} and \cite{DiUj2017} models are consistent with their 
stated uncertainties (and \cite{KaKy2016} perhaps even better than 
that), and that there are significant gains in parameter estimation to 
be made when these uncertainties decrease. We hypothesize that for 
updated simulations using more detailed microphysical descriptions, 
in particular a better treatment of weak interaction, i.e.~neutrino physics, 
it would also be possible to produce analytic models for the results of NR 
and radiative transfer simulations. With a model that both describes the 
improved simulations and has smaller inherent uncertainties in hand, it 
is in principle possible to make precision measurements of ejecta mass 
with results limited only by observation.

To improve the parameter estimation and allow for an extraction of the 
binary properties, we introduced a quasi-universal relations between the 
quotient of baryonic and gravitational mass $M^*/M$ and the compactness 
$C$ of a single neutron star. This relation reduced the number of free 
parameters for the parameter estimation and consequently improved the 
extraction of the individual binary parameters. We also compared the 
parametrized models with other kilonova models and lightcurves of other 
transients. As expected, the lightcurves of a blue kilonovae precursor 
and a SN Ia cannot be approximated by the models, which shows that the 
parametrized models could also be used to rule out some of the possible 
measured transients. We also found that other kilonovae lightcurves, 
\cite{BaKa2016} and \cite{RoFe2017}, are not accurately described as 
well. This is caused by the difference in the underlying radiative 
transfer simulations on which the models are built, which emphasizes 
again the need to improve and update kilonova models in the future.

We also showed how to include the posterior samples from GW signals from 
a binary-neutron star or black hole-neutron star to give further 
constraints on parameters for the lightcurves. We showed improved 
constraints on the ejecta properties $M_{\rm ej},v_{\rm ej}$ and the 
binary parameters $M_{\rm c},q$ using a combination of GW and EM 
observations. This motivates combined analysis in the case of a 
kilonovae detection coincident with a GW trigger.

However, a number of hurdles remain. Mostly due to the large 
uncertainties in the ejecta mass, velocity, density profile, the effect 
of thermal efficiency, and the estimated opacity in the ejected 
material, there are large biases and parameter estimation with the 
current existing models is hampered. To overcome these issues, 
improvements have to be made in numerical relativity by performing 
longer simulations which include additional physics such as other ejecta 
components from magnetic driven winds or neutrino outflows, see 
e.g.~\cite{SuEA2008,MeQa2008,DeOt2008,PeRo2014}. Additionally, improved 
radiative simulations will be needed. Based on those simulations, new 
parametrized models could be developed in the future.

For future application, it will also be useful to consider how to implement a 
search strategy in existing data sets when 
the lightcurves are not necessarily well sampled. 
This would optimize the tiling and time allocation strategies of 
existing searches for GW counterparts with telescopes with wide fields of view.\\

A code to produce the results in this paper is available at 
\url{https://github.com/mcoughlin/gwemlightcurves} for public download.
Required for analysis are text files of lightcurves from models of 
interest in magnitudes, typically available from groups developing kilonova models. 
Furthermore, the kilonovae model of \cite{KaKy2016} can be found online on 
\url{www2.yukawa.kyoto-u.ac.jp/~kyohei.kawaguchi/kn\_calc/main.html}
and the model of~\cite{DiUj2017} on \url{www.aei.mpg.de/~tdietrich/kn/main.html}. 

\section{Acknowledgments}

The authors would like to thank Zoheyr Doctor for a careful reading of an earlier version of the manuscript.
MC is supported by National Science Foundation Graduate Research 
Fellowship Program, under NSF grant number DGE 1144152. CWS is grateful 
to the DOE Office of Science for their support under award DE-SC0007881. 
MU is supported by Funda\c{c}\~ao de Amparo \`a Pesquisa do Estado de 
S\~ao Paulo (FAPESP) under the process 2017/02139-7. KK is supported 
by JSPS Postdoctoral Fellowships for Research Abroad.

\bibliographystyle{aasjournal}
\bibliography{references}

\end{document}